\documentclass[times]{TRR}
\usepackage{tabulary,xcolor}
\usepackage{tcolorbox}
\usepackage{enumitem}
\usepackage{amsfonts,amsmath,amssymb}
\usepackage[T1]{fontenc}
\usepackage{longtable}
\usepackage{amsfonts,amsmath,amssymb}
\usepackage{moreverb,url}
\usepackage{lscape}

\def\tblbottomrule{\noalign{\vspace*{6pt}}\hline\noalign{\vspace*{2pt}}}

\usepackage[colorlinks,bookmarksopen,bookmarksnumbered,citecolor=red,urlcolor=red]{hyperref}

\newcommand\BibTeX{{\rmfamily B\kern-.05em \textsc{i\kern-.025em b}\kern-.08em
T\kern-.1667em\lower.7ex\hbox{E}\kern-.125emX}}

\def\volumeyear{2020}

\begin{document}

\runninghead{Ghalib Ahmed Tahir, Chu Kiong Loo, Nadine Kong and Foong Ming Moy}

\title{A Review of Critical Features and General Issues of Freely Available mHealth Apps For Dietary Assessment}

\author{Ghalib Ahmed Tahir\affilnum{1}, Chu Kiong Loo\affilnum{1},  Nadine Kong \affilnum{2} and Foong Ming Moy \affilnum{2}}

\affiliation{\affilnum{1}Department of Artificial Intelligence, Faculty of Computer Science and Information Technology, University of Malaya, Kuala Lumpur, Malaysia\\
\affilnum{2}Julius Centre University of Malaya, Department of Social and Preventive Medicine, Faculty of Medicine,
University of Malaya, Kuala Lumpur, Malaysia}

\corrauth{}

\begin{abstract}
Obesity is known to lower the quality of life substantially. It is often associated with increased chances of non-communicable diseases such as diabetes, cardiovascular problems, various cancers, etc. Evidence suggests that diet-related mobile applications play a vital role in assisting individuals in making healthier choices and keeping track of food intake. However, due to an abundance of similar applications, it becomes pertinent to evaluate each of them in terms of functionality, usability, and possible design issues to truly determine state-of-the-art solutions for the future. Since these applications involve implementing multiple user requirements and recommendations from different dietitians, the evaluation becomes quite complex. Therefore, this study aims to review existing dietary applications at length to highlight key features and problems that enhance or undermine an application's usability. For this purpose, we have examined the published literature from various scientific databases of the PUBMED, CINAHL (January 2010-December 2019) and Science Direct (2010-2019). We followed PRISMA guidelines, and out of our findings, fifty-six primary studies met our inclusion criteria after identification, screening, eligibility and full-text evaluation. We analyzed 35 apps from the selected studies and extracted the data of each of the identified apps. Most of the apps are engaging, according to user feedback (68\%). 62\% of the apps provide timely alerts to the user, and 53\% of survey apps include goal-settings features.  We indicated existing apps are lagging in several aspects. Only 37\% of the survey application have included validated databases, 22\% of the surveyed applications have addressed data privacy issues, and three applications out of 35 provide offline access to the user. Following our detailed analysis on the comprehensiveness of freely available mHealth applications, we specified potential future research challenges and stated recommendations to help grow clinically accurate diet-related applications.
\end{abstract}

\maketitle
Despite considerable advancements in medicine today, the number of people getting affected by chronic diseases is significantly greater due to unhealthy lifestyles. Obesity is one of the most common contributing factors to chronic diseases, affecting almost every part of the world from middle to lower-income countries. According to a survey in 2016, 1.9 billion adults aged 18 years and older were overweight \cite{1}. The prevalence of the aforementioned diseases poses serious concerns. However, determining the right remedial measures is dependent on different factors ranging from a person's genetics to lifestyle, which need to be adjusted according to the cause and severity of the condition. Treatment may include medication, lifestyle changes \cite{92} such as choosing healthier food alternatives, exercise, and requiring patients to follow a customized diet plan \cite{93}. 

\par On the other hand, with rapid technological advancements and increased usage of handheld devices such as smartphones, tablets, and smartwatches, people's reliance on these devices has undoubtedly grown beyond their utility as a means to communicate. The number of mobile users in 2012 for Android and iOS devices \cite{2} increased from 640 million to 2,562 million in 2016 \cite{3}. These days, smartphone applications are being extended to support electronic healthcare practices \cite{4}, \cite{5}, and evidence across several fields show promising results which support the feasibility, acceptability and efficacy of digital health interventions in different medical conditions. These conditions include but not limited to managing adolescent health and wellness \cite{97}, interventions in sickle cell disease \cite{99}, pediatric cancer \cite{98} \cite{102}, chronic health conditions \cite{100} and improving adherence to preventive behaviour \cite{101}. Similarly, e-Health and related diet-related applications are being increasingly used for professional and personal purposes. People are using these applications to make healthier lifestyle choices. Generally, these apps provide instant nutritional values of food items with barcode scanners, which is extremely helpful for people suffering from non-communicable diseases, and others who intend to choose healthier products \cite{1}. These applications not only assist users with the selection of more nutritious alternatives but also allow them to self-monitor their physical activity and diet intake by using behavioural strategies of goal settings \cite{6}. Moreover, these applications are designed to cater to various age groups, including children battling obesity from a very young age.

In this regard, recent developments in artificial intelligence-based functionalities and hardware capacity enhancement of handheld devices have led to the development of automatic food recognition and calories estimation methods, making them an essential subset of e-health applications. 
\par Regardless of numerous diet-related applications freely available today, scientifically proven guidelines (both in terms of usability and functionality) have not yet emerged from the users' and the dietitian's perspective. Also, the author's first-hand app development experience \cite{7} suggested a dire need to have in-depth knowledge about the state-of-the-art diet-related applications. To develop this understanding, the first step involves identifying key components relevant to existing diet-related applications, which are categorized in terms of general issues faced by dieticians and users, including user experience of both parties and functionalities required by each of them, respectively.
 \par While the development of diet-related applications requires a significant amount of time and effort, general issues like their credibility remain a question. The term "Credibility" here refers to the authenticity or scientific validation of an application to achieve goal during trials.   Another challenge present applications face is the maintenance of an updated food composition database, as new food products are being continuously introduced in the market. Mobile app developers also find it challenging to determine target users, their needs, and potential feedback to improve functionality and usability of apps \cite{7} \cite{8}. Thus, an application with a good user experience may increase its preference over others despite offering lesser functionality. Therefore, the development of such applications should strongly consider essential factors like usability and 'ease of use' \cite{9}  \cite{10},  as poor usability can result in users switching over to alternative options \cite{11} \cite{7} \cite{12}. 
\par The following paper aims to provide a review of existing diet-related applications and seeks to equip researchers and dietitians with comprehensive knowledge about general issues encountered by their users in terms of usability and functionality. Thereby laying the foundation for developing state-of-the-art generalized solutions that can cater to vastly varying user needs. Moreover, other fields in which evidence supports the effectiveness of digital interventions can learn a valuable lesson from the findings of this study.

\section{Methods}
We have developed the review protocol by defining our research questions and considering multiple inclusion/exclusion criteria. 
Then we formally defined our search strategy by identifying the search terms and carried out the search using the electronic database of PUBMED, CINAHIL, and Science Direct.
Following this,  we selected relevant studies based on our study selection criteria. Then we extracted the data and presented our results. 
\subsection{Research Questions}
The primary aim of this review is to answer the research questions shown in Table \ref{tw-ea59a9ad71c5}.\\
\begin{table*}[!ht]
\caption{{Research Questions} }
\label{tw-ea59a9ad71c5}
\def\arraystretch{1}
\ignorespaces 
\centering 
\begin{tabulary}{\linewidth}{p{\dimexpr.16200000000000003\linewidth-2\tabcolsep}p{\dimexpr.43160000000000004\linewidth-2\tabcolsep}p{\dimexpr.4064\linewidth-2\tabcolsep}}
\hline No & Research question & Motivation\\
\hline 
RQ1 &
  What are the general problems that are resolved by the freely available diet-related applications? &
  To provide information about the general problems faced by dietary assessment apps such as frequent app crashes \cite{87}, cumbersome process of entering meal information, demotivating information displays \cite{54}, periodic notifications, difficulty in portion size estimation \cite{85}, credibility \cite{83,89}, etc.\\
RQ2 &
  What are back-end application issues resolved by the freely available dietary applications? &
 To provide information about the application's stability, usage reports \cite{109}, data confidentiality \cite{90}, and offline accessibility-related issues faced by diet-related applications.\\
RQ3 &
  To what extent do the freely available dietary applications fulfill user interface requirements? &
  To provide information regarding the critical user interface components \cite{28,81,82,79,80} catered by diet-related applications.\\
RQ4 &
   What are the dietary components and critical features implemented by the freely available dietary applications? &
  To determine and provide information regarding the dietary components  \cite{83,84,89} and critical features implemented by existing diet-related applications.\\
RQ5 &
  What are the benefits and challenges stemming from the included case studies? &
  To summarize the benefits of dietary-related apps and the challenges they face based on the included studies. \mbox{}\protect\newline \\
\hline 
\end{tabulary}\par 
\end{table*}
\subsection{Inclusion/Exclusion Criteria}
The studies that met all of the following criteria are selected for this review.
\\ IC1. Papers related to dietary applications for smartphones (iPhones, Android phones, and Blackberries) and modern commercially available portable devices such as iPads and Personal Digital assistants (PDAs).
\\ IC2. Content is written in the English language only.
\\ IC3. The study must be a full peer-reviewed paper (not an abstract).
\\ IC4. Dates of Publication: PubMed and CINAHL: 1/1/10-31/12/19, Science Direct: 2010-2019
\par This review excludes the following studies that are conformed to at least one of the following criteria. 
\\ EC1. Studies without a clear description of dietary application mentioned.
\\ EC2. Dietary applications that are not freely available.
\subsection{Database Identification}
We choose PubMed, CINAHL and Science Direct due to the following reasoning. PubMed database gives a publicly available search interface for MEDLINE and National Library of Medicine, which makes it one of the most widely accessible biomedical resources globally \cite{94}. Similarly, the CINAHL database provides allied health care literature, thus making it a good resource for literature related to mHealth applications \cite{95}. We selected Science Direct as it provides broad access to a database of scientific and medical research \cite{96}. 
\subsection{Search Strategy}
\begin{figure}[!ht]
\centering
\includegraphics[scale=0.35]{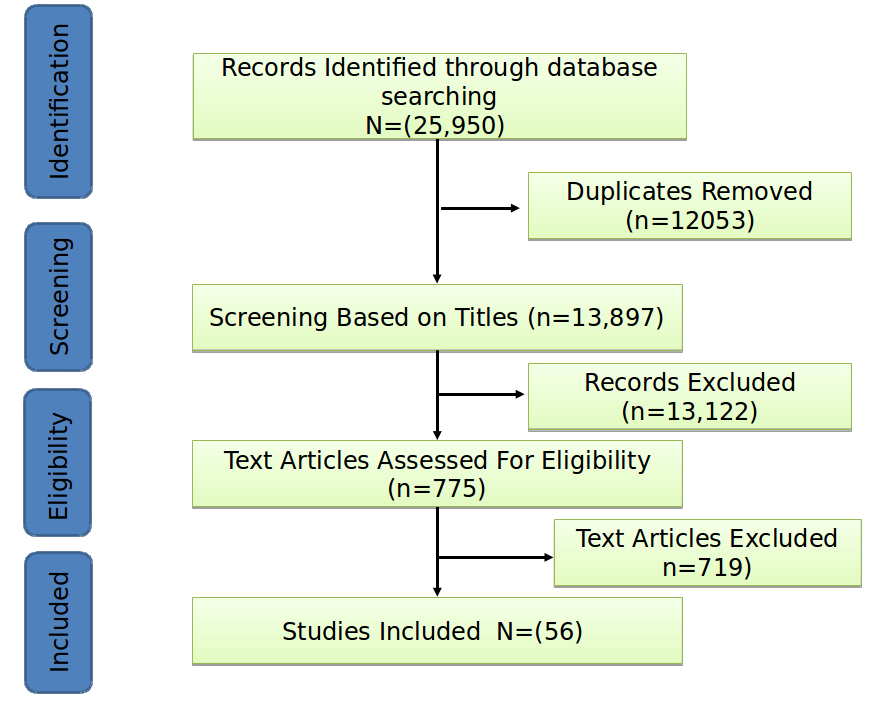}
\caption{PRISMA flow chart of identification, screening, eligibility and inclusion of studies.}
\label{prismaflowchart}
\end{figure}

We have carefully defined the search terms based on initial screening through a consensus among authors to investigate the diet-related mobile applications. Terms such as cellular phone, mobile phone, smartphone, mHealth, iPads combined with terms like diet, food and nutrition are qualified as keywords in our work.
In PubMed, we limited the search to research articles of Clinical Trial, Meta-Analysis and Randomized Controlled Trial published between 1st January 2010 and 31st December 2019. For the CINAHL database, we limited the search to full-text research articles published from 1st January 2010 to 31st December 2019. For Science Direct, we defined the search to research articles published between \textbf{2010} and \textbf{December 2019}.
Search targets the following keywords ("cellular phone" AND diet, "mobile phone" AND diet, smartphone AND diet, mHealth AND diet, iPads AND diet,  "cellular phone" AND food, "mobile phone" AND food, smartphone AND  food, mHealth AND food, iPads  AND food, "cellular phone" AND nutrition, "mobile phone" AND nutrition, smartphone AND nutrition, mHealth AND nutrition, iPads AND nutrition ). The Boolean AND joins the two major parts. These yield 25950 results,  reduced to 13,897 after duplicates removal. They are screened based on titles, and we accessed a total of 775 articles for eligibility against our inclusion/exclusion criteria and study selection process. We scanned references of eligible studies to identify additional studies, but we have not included additional studies in this review. Finally, we included a total of 56 studies in this review process. Table \ref{tw-4cde70680f7f} shows the search terms and their corresponding search results. The PRISMA diagram in Figure \ref{prismaflowchart} shows the search flow and inclusion/exclusion of studies.
\begin{table*}[!ht]
\caption{{Search Results (The Boolean AND joins the two major parts)} }
\label{tw-4cde70680f7f}
\def\arraystretch{1}
\ignorespaces 
\centering 
\begin{tabulary}{\linewidth}{p{\dimexpr.40760000000000005\linewidth-2\tabcolsep}p{\dimexpr.194\linewidth-2\tabcolsep}p{\dimexpr.1919\linewidth-2\tabcolsep}p{\dimexpr.20650000000000002\linewidth-2\tabcolsep}}
\hline Search Strings & \multicolumn{3}{p{\dimexpr(.5924000000000001\linewidth-2\tabcolsep)}}{Search Results}\\
\hline 
 &
  PUBMED &
  CINAHL &
  Science Direct\\
"Cellular Phone" AND Diet &
  5 &
  104 &
  87\\
"Mobile Phone" AND Diet &
  51 &
  622 &
  1,057\\
"Mobile Telephone" AND Diet &
  2 &
  106 &
  79\\
Smartphone AND Diet &
  89 &
  909 &
  1,071\\
mHealth AND Diet &
  211 &
  573 &
  200\\
iPads AND Diet &
  6 &
  287 &
  263\\
"Cellular Phone" AND Food&
  1 &
  180 &
  400\\
"Mobile Phone" AND Food&
  30 &
  1,018 &
  4457\\
"Mobile Telephone" AND Food&
  1 &
  132 &
  314\\
Smartphone AND Food&
  64 &
  1,396 &
  3,418\\
mHealth AND Food &
  103 &
  640 &
  239\\
iPads AND Food &
  5 &
  610 &
  843\\
  "Cellular Phone" AND Nutrition &
  2 &
  136 &
  106\\
"Mobile Phone" AND Nutrition &
  41 &
  757 &
  1,191\\
"Mobile Telephone" AND Nutrition &
  2 &
  121 &
  90\\
Smartphone AND Nutrition &
  90 &
  1,030 &
  1,139\\
mHealth AND Nutrition &
  183 &
  663 &
  171\\
iPads AND Nutrition &
  5 &
  344 &
  316\\
\hline 
\end{tabulary}\par 
\end{table*}
\subsection{Evaluation Criteria and Data Extraction}
Reviewer GAT extracted all the selected studies' key characteristics  (study population, location, mobile app details, and aim of the survey) shown in Table \ref{tw-b3e1fe13b830}. Similarly, the Expert Group compromised of the authors of this manuscript identified the attributes of each research question shown in Table \ref{tw-6aed7cb5ec22}, \ref{tw-8cd9fe71280b}, \ref{tw-2e7650d47b96} and \ref{tw-e4bf42e0a922} that are mentioned in the existing literature of mHealth apps for data extraction to answer our questions. 
Under the heading of general issues, we assessed the difficulty in portion size estimation \cite{84}, demotivating information, dependence on expensive electrical devices such as fit brands \cite{54}, the credibility of the database \cite{83,89} etc. Table S1 and S2 provide the extracted data of general issues in the supplementary material. We extracted the data regarding stability of the application \cite{87}, usage reports \cite{109}, data confidentiality \cite{90} and offline accessibility \cite{110} for back-end application issues. Table S3 provides the extracted data in the supplementary material. For user interface requirements \cite{28,81,82,79,80}, we extracted data of the attributes mentioned in Table \ref{tw-2e7650d47b96}. Table S4 and S5 in the supplementary material provides the extracted data. Under the heading of dietary components \cite{83,89}, we extracted the details of the attributes shown in Table \ref{tw-e4bf42e0a922}. Table S6 and S7 in the supplementary material provides the extracted data.
We carried out the whole process by completing the data extraction forms. Two researchers verified the data's soundness and ensured data extracted from each study justified the study's aim. When the publications identified in the searches did not provide sufficient detail of mHealth apps, additional literature, websites, contacts with authors, or application use itself was used to fill gaps.

\section{Results} 
This section presents the results of the essential characteristics of each selected study. It shows the results obtained from the extracted information to answer our research question. The brief detail of all the data extracted to answer our research question is provided in the study's supplementary material.
\subsection{Study Selection}
The database search yielded 25,950 results. After removing duplicates, we screened 13,897 based on titles. Out of that, we excluded 13,122  studies, and 775 article texts were assessed for eligibility by reviewers. Finally, we included 56 studies after excluding 719 text articles. The first study included is from 2010. From 2013 to 2017, the publication rate increased by 55.08\%, with the highest number of studies published studies in 2017 (22.3\%). We have the publication year of all the included studies in our key characteristics table.
\par In the subsequent sections, we have briefly described mobile applications' status as per our research questions. We have evaluated existing dietary applications by keeping in view critical features and general issues mentioned by dietitians and users and supported by existing literature of mHealth apps.

\subsection{ RQ1: What are the general problems that are resolved by the freely available diet-related applications?}
\begin{figure}[!htbp]
\centering
\includegraphics[scale=0.6]{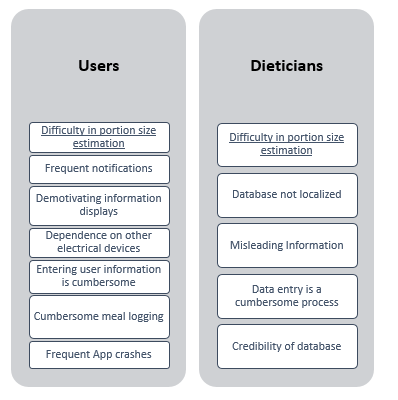}
\caption{General issues faced by freely available diet-related applications }
\label{issuesofusers}
\end{figure}

To determine the general issues found in existing applications, we have categorized important parameters from different perspectives of users and dietitians as shown in Figure \ref{issuesofusers}. For this purpose, we surveyed 35 freely available mHealth apps from included studies. 
\par
\begin{figure*}[!htbp]
\centering
\includegraphics[scale=0.4]{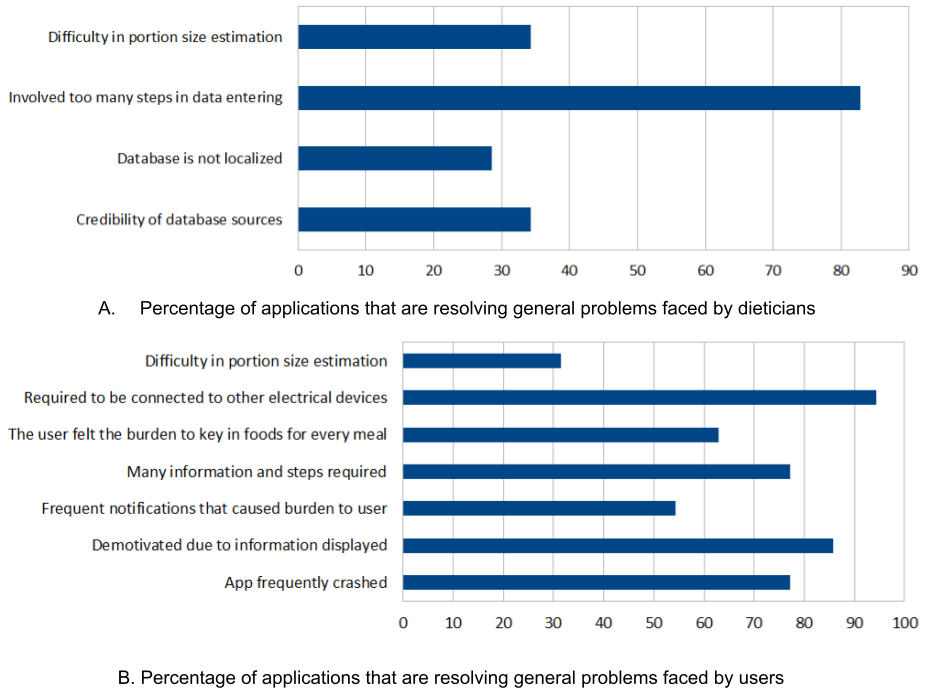}
\caption{(A) and (B) shows percentage of applications that are resolving general problems faced by dietitians and users}
\label{generalproblemsbyusersdieticians}
\end{figure*}

 These general issues mentioned by dietitians and users in existing mHealth apps include credibility, \cite{83} localization of database sources, \cite{85}, and difficulty in portion size estimation \cite{84}. Moreover, applications that require users to go through multiple steps for data entry \cite{86} make the process cumbersome and negatively impact the whole user experience. (Figure \ref{generalproblemsbyusersdieticians}) (A) shows the percentage breakdown of applications that looked to resolve these issues. 
\par Credibility of database sources is one of the major reasons for dietitians to not recommend apps to clients or patients due to concerns regarding their validity and questionable feedback in terms of accuracy. Nearly 34\% of the applications \cite{21,22,25,29,35,37,47,52,57,58,64,66,68} managed to resolve this issue by providing extensive details about the database, especially in terms of its sourcing.  However, there are still some applications that offer little or no information about their database sources \cite{18,20,27,31,33,73,38,49,74,75,63}. Remaining applications \cite{16,19,23,28,6,32,36,43,44,69,71} did not address the issue of credibility of database resources. On other hand only 10 out of 35 \cite{21,22,6,37,43,47,49,52,58,63}  applications surveyed had localized databases which are specific to certain region or culture.
\par Subsequently, portion size estimation also plays a vital role when it comes to dietary applications. This component usually requires prior contextual knowledge to ensure better accuracy. Several apps deal with fixed food measurements in terms of serving size, weight, or other simple household measurements. Generally, it is hard for most people to convert what they see on their plates to these measurements for entering into dietary apps. Moreover, when it comes to Asian food, estimating portion size becomes even more challenging when multiple food items are mixed or placed on top of each other. Therefore, over or underestimating portion size is common for unskilled individuals, even more so in Asian families where each meal consists of multiple side dishes. These challenges make the estimation of portion size a complicated task for machine learning researchers, application developers, and dietitians. Many existing applications \cite{21,23,27,28,29,6,32,73,35,36,70,52,71} do not accommodate features which support estimation of portion size. Alternately, many existing applications \cite{16,18,19,20,21,23,27,28,29,32,33,35,70,73,37,38,43,49,74,75,57,58,63,64,66,69,71} require minimum steps for data recording ensuring a smooth user experience.
\par General issues faced by users involve frequent app crashes \cite{87}, cumbersome process of entering meal details, demotivating information displays, dependence on expensive electrical devices such as fit bands  \cite{54}, frequent notifications and difficulty in estimating portion size \cite{85}. As shown in figure \ref{generalproblemsbyusersdieticians} (B), out of 35 surveyed applications, 6 applications \cite{43,44,47,49,74,75} had no information about application's stability in terms of app crashes. Both "happy" \cite{18} and "lose it" \cite{19} experienced frequent app crashes. The remaining applications do not experience frequent app crashes.
\par Furthermore, large number of surveyed applications display motivating information and require less steps to record user information \cite{16,18,19,20,21,22,23,27,28,29,32,33,35,36,37,73,38,43,74,57,58,63,64,68,66,69,71}. Frequent notifications often bother the user; especially when the app provides notifications after nearly every step. Also, 40\% of surveyed applications had no information available about notification or reminder settings, \cite{21,25,27,6,31,32,73,36,37,44,47,52,75,64} whereas remaining applications accommodated notification system except “lose it” \cite{19}. Furthermore, all applications except FoodWiz2 \cite{66} and MyFitness Pal (Log2Lose) \cite{71} do not require any sort of electrical device. Also, almost 37\% of applications \cite{20} \cite{22} \cite{25} \cite{27} \cite{6} \cite{31} \cite{32} \cite{47} \cite{52} failed to address the cumbersome process of entering meal information. In addition to this, a large number of existing applications do not address the difficulty in portion size estimation. This is due to multidimensional challenges, from users’ perspective; as it is difficult to estimate the portion size of food items used during preparation of food at different restaurants. Moreover, lack of guidance/reference regarding the quantity, further complicates the procedure of estimating portion size.
\subsection{RQ2: What are back-end application issues resolved by the  freely available dietary applications?}
\begin{figure}[!htbp]
\centering
\includegraphics[scale=0.5]{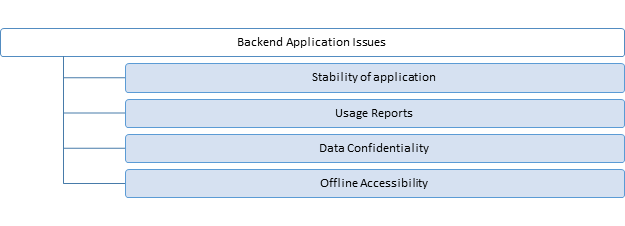}
\caption{Backend application issues}
\label{beapplicationissues}
\end{figure}

The backend is an essential part of any mobile application, as it involves data storage, business logic, and security. Therefore, it plays the role of a server for mobile applications and stores information invisible to the end-users. Figure \ref{beapplicationissues} presents backend application issues faced by end-users. Backend issues generally involve no offline access to the key features \cite{110}, absence of usage reports \cite{109}, privacy or data confidentiality concerns \cite{90}, and frequent application crashes \cite{87}. Therefore, the issues mentioned above should be addressed by mobile applications to enhance the usability or user-friendliness of any diet-related application.
\par Reduced dependence on the Internet will not only improve usability. It will also enhance the app’s responsiveness and data processing, as the end-user is not restricted from recording their data offline. Thus far, only 3 out of 35 surveyed applications allow offline accessibility \cite{23} \cite{43} \cite{71}, whereas other applications require an internet connection for data transmission to their respected servers. Moreover, apps like Diet Cam rely on the client-server configuration for connectivity between mobile phones and databases \cite{22}, which again requires a stable internet connection.
\par Similarly, technical bugs and frequent app crashes resulted in unstable applications \cite{18} \cite{19}\cite{20}\cite{27}\cite{6}\cite{73}\cite{57}\cite{64}. Many of the applications surveyed suffer from technical glitches and slow processing speed due to their dependence on a reliable internet connection. Therefore, offline accessibility can help to address all these concerns.
\par
\begin{figure*}[!htbp]
\centering
\includegraphics[scale=0.4]{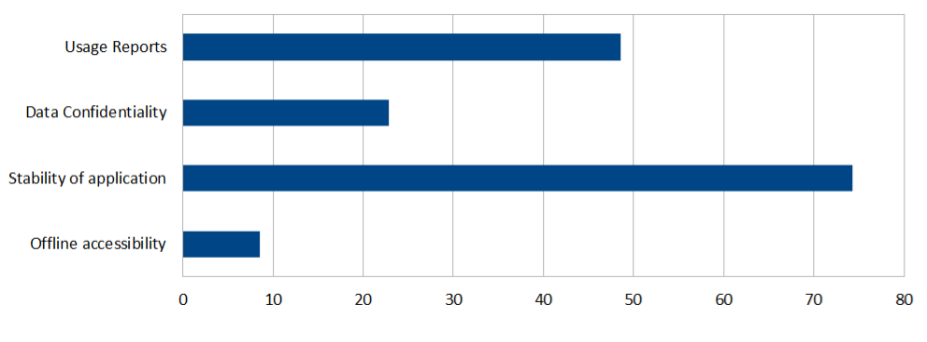}
\caption{Shows percentage of applications that are resolving Backend application issues}
\label{percbackendapplicationissues}
\end{figure*}

 Another important issue is lack of data confidentiality and privacy. Mobile applications, especially health-related applications, should have concrete measures to ensure user's records' confidentiality. Similarly, web services included in mobile applications should extract data without any leakages and minimal pilferage instances. Almost 22\% of surveyed applications offer data privacy, while many studies do not mention this problem. Moreover, some existing applications such as Social POD \cite{16}, Happy \cite{18}, and 'mDPP' \cite{20} report user engagement or adherence to the app declined over time. Figure \ref{percbackendapplicationissues} describes percentage of applications that managed to resolve aforementioned issues.
\subsection{RQ3: To what extent do the freely available dietary applications fulfill user interface requirements?}
Generally, user interface requirements encompass application design, user-friendliness, tutorial page, and two-user dashboard. As application design is one of the main requirements \cite{28,81,82}, as it should be simple, have nice and appropriate icons along with clear font size and color to improve usability.  Out of 35 surveyed applications, 29 applications met the design criteria according to users’ requirements. However, remaining applications \cite{16,20,29,36,38,52} did not endorse their design details.
\par Another feature which enhances usability is the presence of appropriate tutorial pages \cite{79,80}. For many diet-related applications, tutorial pages are preferred to show the metric measurements of serving size of food for a better understanding of the user.  Unfortunately, most of the surveyed apps fail to provide this information, and only 20\% of surveyed applications \cite{21,23,49,58,74,68,71} were able to provide the details for a tutorial page. 
\par Two-user dashboards are another essential feature whereby a simplified or easy-to-use version of the dashboard is available for patients or the general public. A more detailed version is available for dietitians or researchers. Most surveyed applications do not incorporate this feature except for the Dietary Intake Assessment app \cite{51}.
\par Apart from these, user-friendliness has been qualified as the most important UI design component \cite{82}. Applications are considered user-friendly when they have a complete data set, require fewer data entry steps, provide meaningful information, and have a user-friendly interface. Therefore, bugs, glitches, and a cumbersome user interface of apps can negatively influence the app's usability. According to dietitians, almost 77\% of the surveyed applications \cite{18,19,21,22,23,25,28,6,32,35,36,37,43,44,47,49,52,57,58,63,64,66,68,69,71,74,75} were user friendly and had higher rate of user engagement due to presence of simple user interface and interactive design.  According to the users, 91\% of the total applications surveyed were user-friendly. For application design, most of the users prefer simple design and easy-to-use apps.  70\% of applications \cite{37,18,19,21,25,27,28,31,32,73,43,44,49}  were attractive  according to users’ requirements.
\par User interface requirements also involve information to include in the user profile, and notification alerts to the user \cite{81}. Besides basic information, the application should allow users to set goals in terms of desired body weight, and diet \cite{81,83}. Only 53\% of surveyed applications \cite{16,19,24,28,30,33,74,36} included goal-setting feature, while others did not even provide personalized profiles. 
\par
\begin{figure*}[!htbp]
\centering
\includegraphics[scale=0.4]{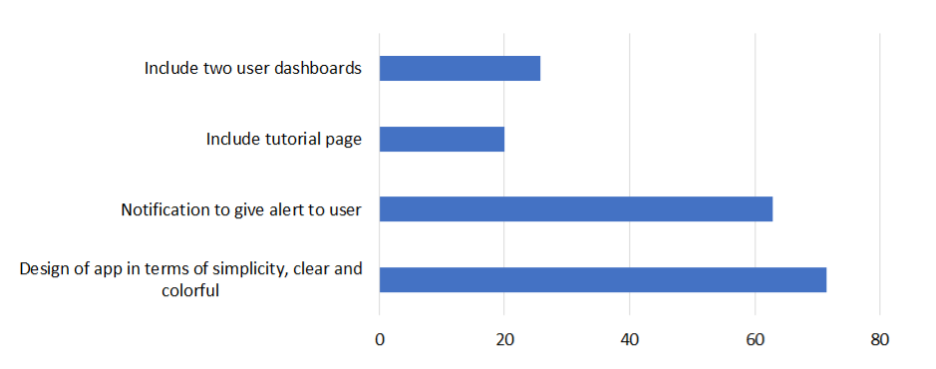}
\caption{Shows percentage of applications that are implementing key user interface features}
\label{percentuserinterface}
\end{figure*}

 Finally, another important feature of diet-related applications is notification alert to users \cite{83}.  Reminders assist users in punching in their updated information and updating their body weight regularly in the app, thereby keeping track of their progress. Similarly, alerts to consume meals at specified times, alerts for calories, drinking water, and doing exercises improve user engagement. 62\% of total surveyed applications \cite{17,19,20,21,24,29,6,32,33,36,43,38,47,49,52,74} were found out to provide such alerts to their users. 
Figure \ref{percentuserinterface} shows the percentage of existing applications that are implementing these features.
\subsection{RQ4: What are the dietary components and critical features implemented by the freely available dietary applications?}
\par Dietary component functionality mainly included evaluation of diet quality, options to add supplements to the diet, history tracking, and storage of these records.  Other essential features include validity and comprehensiveness of database \cite{83,89}, portion size estimation \cite{84} and diet/nutrient summary that provides information in terms of calories for each meal as show in Figure \ref{dietarycomponentapplication}.
Figure \ref{percdietarycomponents} below illustrates the summary of the results gathered from surveyed applications. To provide a good evaluation of diet quality, a dietary app should display macronutrients' balance and include reference values for interpretation. Based on this information, 34 out of 35 surveyed applications were able to assess diet quality properly. 
\par Moreover, options to track users' weekly diet records and their storage on websites for later use are considered important factors that can improve user experience. As per our survey total 26 applications\cite{16,18,19,21,22,23,28,6,29,31,32,73,37,47,49,52,74,57,58,63,64,68,69,71} facilitate users by giving them access to their previous records. Items included in existing applications should also be considered an essential feature, as some applications are particular about specific food items (like beverages). In contrast, other diet apps provide users with options to customize the food choices accordingly.
\par Another important feature which most of applications \cite{16,18,19,20,23,27,28,31,25,35,43,74,38}  (21 out of 35 applications) failed to include, is the incorporation of a reliable and comprehensive food database. Apart from inclusion of database, validity of database also matters and unfortunately only 37\% of the surveyed applications \cite{21,22,25,29,6,35,37,47,52,57,58,64,66} possess validated food database. 
\par Furthermore, applications should display the breakdown of nutrient components of the consumed food items. Data should also include the proportion of calories of each meal (eg. Breakfast, lunch, dinner). However, some of the current applications only include calories per meal, whereas other application like ‘Lose it’ \cite{19} provides information about balanced macronutrients. Another important functionality brought forth by existing applications \cite{19,22,23,25,28,29,6,31,25,37,38,44,47,49,52,75} is portion size estimation. Only few surveyed applications \cite{22,25,31,25,37} use camera to estimate volume of a portion size, while most other surveyed apps rely on the standard household measurements \cite{19,22,23,28,29,6,38,44,,47,49}.
Water is considered an essential component of the human body, ensuring the proper functioning of multiple bodily functions. Therefore, it is equally as important to track users’ water intake. However, 14 out of 35 applications including Happy \cite{18}, Lose it \cite{19}, Metabolic Diet app \cite{21}, MyFitnessPal \cite{23}, and My Meal Mate \cite{6}  allows users to record total water intake. Whereas other diet-related applications tend to miss out on this important feature.
\par As for the nutrient summary, most application databases include calorie information while other surveyed applications provide more specific nutrient information in databases. Apps like “Lose it” \cite{19} provide information about three significant macronutrients like carbohydrates, proteins, and fats, which guide users to make better food choices.
\par To enhance the user-friendliness of an application, tailored messages, feedback, and notifications according to user dietary intake \cite{81} are essential factors to be considered. Acting as guidelines for users, they improve the user-friendliness and user experience of an application. For instance, apps like My Meal Mate \cite{6}, ‘iDAT’ \cite{32}, and MyFitnessPal \cite{23} display remaining calorie allowance to guide the user to achieve dietary goals.
\par Furthermore, applications can make use of visual aids by providing summaries of energy and nutrients intake in the form of a diary, pie chart, table, and progress bar \cite{82} for better comprehension. Applications such as ENGAGED \cite{35}, provide goal thermometers to display user goals and the actual amount of calories and fat (in grams) consumed.
\par
\begin{figure*}[!htbp]
\centering
\includegraphics[scale=0.5]{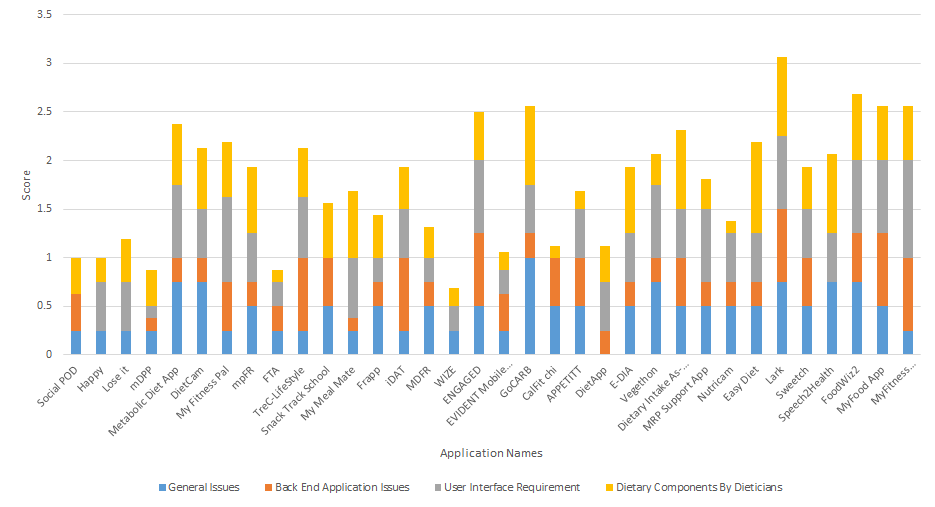}
\caption{Application score by keeping in view requirements from dietician perspective.Equal weight is given to each category and applications fulfilling more requirements have the highest score. The maximum total score is 4, and the max score of each category is one.}
\label{userapplicationrating}
\end{figure*}

\begin{figure*}[!htbp]
\centering
\includegraphics[scale=0.5]{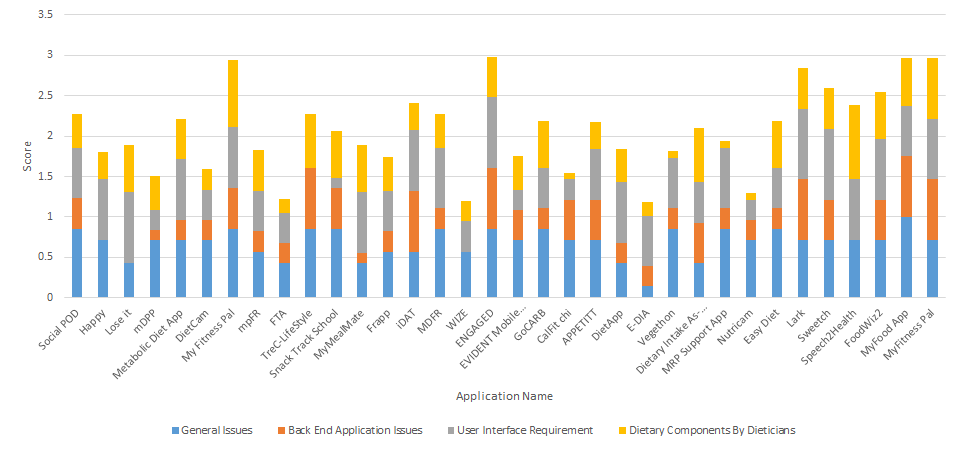}
\caption{Application score by keeping in view requirements from user perspective.Equal weight is given to each category and applications fulfilling more requirements have the highest score. The maximum total score is 4, and the max score of each category is one.}
\label{dieticianapplicatinrating}
\end{figure*}

 Finally, we have ranked these applications based on the number of features they have implemented and mentioned in their study.  Lark application, Food wiz2, Gocarb application had a higher score in fulfilling the number of requirements from dietitians’ or researchers’ reference frames. On the other hand, MyFitnessPal, Engaged, and MyFoodApp focuses more on the general population’s requirements. Figure \ref{dieticianapplicatinrating} and Figure \ref{userapplicationrating} show the applications score.
 \begin{figure}[!htbp]
\centering
\includegraphics[scale=0.6]{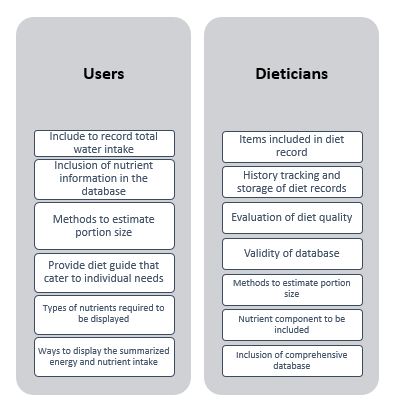}
\caption{Important dietary component by users and dietitians}
\label{dietarycomponentapplication}
\end{figure}

 \begin{figure*}[!htbp]
\centering
\includegraphics[scale=0.4]{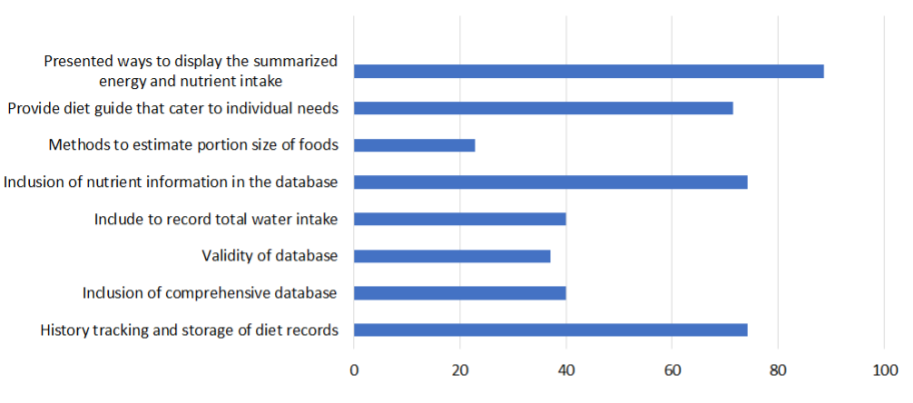}
\caption{Shows percentage of applications that are implementing important dietary components}
\label{percdietarycomponents}
\end{figure*}

\subsection{RQ5: What are the benefits and challenges stemming from the included case studies?}
Current case studies have several benefits that can help users monitor their daily diet and help them resolve diet-related issues. Based on the shortlisted studies, we investigated freely available diet-related applications in terms of features, general problems, and usability challenges. Our findings aim to provide a broader view of current solutions to dietitians, health experts, and researchers alike. Overall, the case studies equip both general users and health experts with information on the critical features that are not catered by most of the existing diet-related apps. Moreover, they will help users quickly determine the viability of existing solutions to recommend further or use the solution that fulfills their needs in the best possible way.  Thus, these studies have paved the way for the research community to introduce standard guidelines for future diet-related apps according to criteria. As a result, the apps will be more substantial for patients, general users, and dietitians.
\par Apart from this, the studies also highlight different challenges that can undermine current applications' actual purpose. Major obstacles include integration and updating large food databases as food recipes, and their nutrient content varies from region to region. In addition to this, new food items are being introduced in the market every day. Therefore, making the design and implementation of such systems a difficult task. Similarly, a database's comprehensiveness is one of the primary requisites of users to keep track of their micronutrients and macronutrients. Other than this, the incorporation of user-friendliness along with notifications and personalized alerts are the challenges that the research community should consider. Also, data security is of vital importance \cite{90} due to strict policies of regulating authorities and rising public concerns over sharing private data. Therefore, diet-related apps should ensure data privacy and confidentiality, which, unfortunately, many surveyed apps fail to address.
\section{Discussion}
We initiated this evaluation because of the rapid recent emergence of freely available diet-related apps coupled with increasing concerns over general issues, usability challenges and missingness of critical features. Following that, we investigated the strength and weaknesses of freely available diet-related apps. The primary emphasis of previous reviews by Rusin et al. \cite{76}, Kankanhalli et al.  \cite{77}, and Prgomet et al. \cite{78} was on functionalities and input methods or the combined intervention of sleep and diet. Similarly, Prgomet et al. \cite{78} focused only on the inclusion of nutrition information in the meal ordering system.
\par We focused on mHealth apps identified from existing publications between 2010-2019.  After carrying out the literature search on three scientific databases, we evaluated 35 mHealth applications based on their usability, critical features and shows their strengths and weaknesses.  The user-friendliness and high engagement are of considerable importance \cite{82}, especially since 68\% of the existing mHealth apps incorporate this feature. We recommend that user's input in the development of mHealth interventions and other considerations for end-users should be sought early on in the process of app or digital health intervention design to ensure long and short term engagement \cite{105} \cite{106} \cite{107} \cite{108}. Similarly, the user notifications are equally important, as it keeps them engaged and motivated \cite{83}. According to the survey, we found that 62\% of the apps provides timely alerts to the user. Likewise, goal setting also holds critical significance, as it gives information about user's personal preferences required for modification of their behaviour accordingly \cite{81,83}.  Therefore, about 53\% of the surveyed applications include the goal settings feature.  
\par Likewise, our findings indicate that existing applications are lagging in various aspects. Despite the importance of the credibility of database resources \cite{83}, only 30\% of surveyed applications highlighted this issue. Besides credibility, the comprehensive validation of the database with detailed information on macronutrients and micronutrients is also essential for clinical use.  However, only 37\% of the applications have included validated databases. Despite the rise of artificial intelligence, the methods for estimating portion size and logging food photos from the camera have made significant advancements \cite{39,55}, many apps still depend on household measurements for portion size estimation or manual entries of the food log. 
\par Due to rising concerns of data security among users \cite{90}, diet-related apps must encrypt user data and use standardized protocols to ensure data privacy and confidentiality. Yet, the results indicate that only 22\% of surveyed applications have addressed data privacy issues.
Similarly, there is a lack of economic data in existing studies to support using mHealth apps for dietary assessment. Although the economic evaluation of mHealth apps is necessary to provide an evidence-based assessment of sustainability and benefits of investing in such technologies. \cite{103} \cite{104}. 
\par Despite considerations that existing diet-related apps should address, all of the studies are valuable to broaden the research community's knowledge. The identified applications in these works serve as a guide for users to choose between healthier alternatives and improve their dietary habits in the long term. 
\par Finally, we have made the following recommendations for the research community based on our study. A localized database is essential for nutritional assessment apps due to variations in the food recipes and diet preferences among different cultures. Future diet-related apps should also consider the technological advancement in artificial intelligence and explore the current methods of logging food and automated portion size estimation from food photos. It is noteworthy that several studies have implemented AI-based strategies, but further investigation of these methods is required on a large scale. Furthermore, there is a dire need to develop standard guidelines for the development of diet-related apps, as standardized solutions will be more reliable in the future for patients, general users, and dietitians. Finally, when designing modern diet-related applications, the research community should consider our findings to enhance the usability and completeness of the solution.

\subsection{Limitations of the data gathered for this study}
This review has limitations that require further investigations. Firstly, the analysis was limited to studies published in the searched databases and only written in the English language.  Related articles in other languages were not included. Secondly, this research does not consider demographic information about a particular race or culture while designing the research questions. 
 \section{Conclusion}
 Dietary apps for nutritional assessment are developed to assist users with their diet-related issues or keep track of their dietary intake. Such apps tend to act as guides and enable users to choose healthier alternatives to improve their nutritional habits in the long term. Therefore, due to the vital importance of diet-related apps, this SLR analyzed a wide range of existing literature on mHealth apps from scientific databases of CINAHL, Science Direct, and PUBMED and shortlisted almost 56 studies. We have investigated the apps' comprehensiveness in terms of critical features, general issues, and usability challenges from general users' reference frames. We have further examined the strength and weaknesses of the existing freely available diet-related apps and summarized concerns and gaps for future work. Our findings show that the credibility of database resources, comprehensive information about macronutrients and micronutrients, validation of database, data privacy, use of AI for food logs, and automated portion size estimation from the pictures are foremost challenges. Addressing the challenges mentioned above will improve the usability and comprehensiveness of diet-related apps. Therefore, making them more substantial for patients, general users, and dieticians. 
Moreover, implementing blockchain technology and health standards for data security, exploring recent trends in continual learning for food recognition, and outlining standard guidelines for regulating apps are essential future topics that can be explored.

\begin{acks}
This research was supported by the UM Partnership Grant: Project No: RK012-2019 from University of Malaya, IIRG Grant (IIRG002C-19HWB) from University of Malaya, International Collaboration Fund for project Developmental Cognitive Robot with Continual Lifelong Learning (IF0318M1006) from MESTECC, Malaysia and  ONRG grant (Project No.: ONRG-NICOP- N62909-18-1-2086) / IF017-2018 from Office of Naval and Research Global, UK.
\end{acks}

\clearpage
\begingroup
\onecolumn
\begin{longtable}{p{\dimexpr.2\linewidth-2\tabcolsep}p{\dimexpr.3\linewidth-2\tabcolsep}p{\dimexpr.1\linewidth-2\tabcolsep}p{\dimexpr.4\linewidth-2\tabcolsep}}
\caption{Main characteristics of the selected studies} 
\label{tw-b3e1fe13b830}
\def\arraystretch{1}\\\endfirsthead \hline \noalign{\vskip3pt} \noalign{\textit{Table \thetable\ continued}} \noalign{\vskip3pt} \hline \endhead \hline \noalign{\vskip3pt} \noalign{\textit{\hfill Continued on next page}}  \endfoot \hline \endlastfoot 
\hline Author, Year & Study Population, Country & Mobile App & Aim of the Study\\
\hline 
Hales et al., 2016  \cite{16} &
  N=51 Overweight adults, South Carolina USA \mbox{}\protect\newline  &
  Social POD &
  To scrutinize the efficacy of the mobile app in terms of weight-loss  \mbox{}\protect\newline \\
McCarroll et al., 2015 \cite{17} &
  N=50 overweight participants, USA &
  LoseIt &
  Investigate the usefulness of the lifestyle intervention program designed for weight loss of early-stage obese/overweight cancer survivors in the last three years.\\
Ribeiro et al., 2017 \cite{18}&
  N=32 volunteers, Portuguese &
  Happy &
  To scrutinize the effectiveness of the behavior change approach to persuade users to choose healthier alternatives to minimize the risk of several types of cancer.\\
Burke et al., 2017 \cite{19}&
  N=39 volunteers, Pittsburgh, USA &
  Lose it &
  A randomized clinical trial to examine the benefit of 1{\textendash}4 daily FB messages customized to dietary recordings through a smartphone for weight loss purposes\\
Fukuoka et al., 2015 \cite{20}&
  N=61 overweight adults, San Francisco, USA &
  mDPP &
  Explore the usefulness of a diabetes prevention intervention along with a smartphone application and pedometer among overweight individuals at risk of type 2 diabetes.\\
Ho et al., 2016 \cite{21}&
  N= 5 families using the app, Vancouver, Canada &
  Metabolic diet App &
  To implement apps for medical diet for all intrinsic delusions of amino acid metabolism.\\
Kong and Tan, 2012 \cite{22}\mbox{}\protect\newline  &
N=21 restaurants are covered to verify their vision-based method, USA
   &
  DietCam &
  To monitor the food intake with few human interventions
  \\
Levinson et al., 2017 \cite{23}&
  N=105 participants diagnosed with eating disorder, USA &
  My Fitness Pal &
  To examine the relationship of calorie tracking app with eating disorder pathology\\
Wharton et al., 2014 \cite{24}  &
  N=19 participants, Arizona, USA &
  Lose It &
  To study the effectiveness of smartphone application for self-monitoring of dietary intake and weight loss compared to the traditional approaches of counseling and recording methods.\\
Six et al., 2010 \cite{25} &
  N= 78 participants, USA &
  mpFR &
  To compare the calculated energy and protein content of foods provided by the app with the published estimates in the FNDDS.\\
Gilson et al., 2014  \cite{26}   & N= 44 drivers, Australia
   & UP
   & To study how drivers have accomplished the process from baseline phase to smartphone application for diet monitoring.

  \\
Holmen et al., 2014  \cite{27}&
  N=151 participants, Norway &
  FTA &
  To determine the effectiveness of a smartphone{\textendash}based self-management system to monitor the levels of glycated hemoglobin A1c (HbA1c) and health-related quality of life.\\
Gabrielli et al., 2017   \cite{28}&
  N=6 families participating, Italy &
  TreC-LifeStyle &
  To demonstrate the TreC-LifeStyle nutrition education application's design and development along with the results of a formative analysis of families.\\
Van Lippevelde et al., 2016 \cite{29}&
  N= 1400 participants, Belgium \mbox{}\protect\newline  &
  Snack Track School &
To study the benefits of a dual-system model combined with the behavioral change approach while targeting both the reflective and automatic processes.\\
Carter et al.,2017 \cite{30}&
  N= 43 participants, United Kingdom &
  My Meal Mate &
  To study the association among frequency and pattern of self-monitoring diet through a smartphone app and weight loss.\\

Goh et al., 2015  \cite{32}&
  N= 84 participants, Singapore &
  iDAT &
  To study short-term (8-week) trajectories resulted from iDAT application between patients with type 2 diabetes and determined the patient's behavior associated with every trajectory.\\
Kerr et al., 2012 \cite{33} &
  N=220 participants, Perth Australia &
  CHAT &
  To study the effectiveness of a smartphone app for monitoring dietary intake,  giving feedback through text messages to motivate young individuals for bringing change in their diet through more intake of fruits, vegetables, and less consumption of junk foods.  \mbox{}\protect\newline \\
Mann et al., 2015 \cite{73}& N= 26 participants, Australia
   & WIZE
   & To investigate the usability, content and acceptability of a smartphone application developed to boost the absorption of dietary iron.
  \\
Pellegrini et al., 2012 \cite{35} \mbox{}\protect\newline  &
  N=96 obese adults,USA &
  ENGAGED &
  Examine the efficacy of using a smartphone application for weight loss purposes.\\
Recio-Rodriguez et al., 2016 \cite{36} &
  N=833 individuals,Spain &
  EVIDENT Mobile phone apps &
To compare the expected amount of macro and micronutrients, calorie intake, and misuse of alcohol by using the mobile app to gather data with the data assessed using the food frequency questionnaire.\\
Rhyner et al., 2016 \cite{37} &
  N=19 Participants, Switzerland &
  GoCARB &
  Evaluate the accuracy of GoCARB application used by individuals with type 1 diabetes and collate it to their performance in carbohydrate counting.\\
Seto et al., 2016  \cite{38} &
  N=12, Participants,USA &
  CalFit Chi &
   To investigate the relation between individual-based modeling methods and the eating behavior of a person. Also, to determine the usefulness of such approaches compared to conventional regression models.\\
Six et al., 2011   \cite{25-a} & N=78 participants,Australia &
  mpFR & To compare measured energy and protein content of foods with the published estimates in the FNDDS among adolescents \mbox{}\protect\newline \\
Zhu et al., 2010 \cite{39} & N= 78 participants, USA
   & Food Record App
   
   &
   To automate the food logging process by using image processing methods and reduce the burden on the respondents.
  \\
Probst et al., 2015 \cite{41} &
  No Participants, App Prototype, Australia &
  Prototype App &
  To develop a prototype for automatic food recognition via a use of image processing methods\\
Serrano et al., 2016 \cite{42} \mbox{}\protect\newline  &
  No participants, Health data mining, USA &
  Lose It &
 To scrutinize the data from a commercial dietary app (Lose It!) about subgroups who were successful in the weight loss program by exploratory examinations and confirm the results' stability. \mbox{}\protect\newline \\
Farsj{\o}, et al., 2017 \cite{43} \mbox{}\protect\newline  &
  N = 4 participants, Norway &
  APPETITT &
  To access the effectiveness of apps for inexperienced users of technology and examine its contribution for encouragement and orientation about meals.\\
Torre D\'{\i}ez, I., et al., 2017 \cite{44} \mbox{}\protect\newline  &
  N= 150 participants, Spain & DietApp&
  To implement and study smartphone application to provide suggestions to achieve a healthy diet according to individual's age, clinical history, and physical condition.\\
Carter, M. C., et al., 2013 \cite{6} & N= 9 participants, United Kingdom
   &
  My Meal Mate &
  They introduced and developed a smartphone application based on focus group interviews for weight loss.
  \\
Nystr{\"{o}}m, C. D. et al., 2016  \cite{46} \mbox{}\protect\newline  &
  N= 39 participants, Sweden &
  TECH &
  To evaluate energy intake (EI) using TECH application with total energy expenditure (TEE) calculated using doubly labeled water (DLW), and to compare dietary intakes using TECH app with intakes obtained through 24 h dietary recalls. \mbox{}\protect\newline \\
Rangan, A. M., et al., 2016 \cite{47} &
  N=80 university Students, Australia &
  e-DIA &
 To investigate the validity of the e-DIA app's comparison with the 24-h recall approach to calculate dietary intake of food groups.\\
Casperson, S. L. et al., 2015   \cite{31}&
  N= 18 participants,USA &
  FRapp &
  To determine the amenability of adolescents to use the FRapp to monitor their dietary intake.\\
Mummah, S. et al., 2016 \cite{49} \mbox{}\protect\newline  &
  N=17 participants,USA  \mbox{}\protect\newline  &
  Vegethon &
  To develop a theory-driven mobile application for increasing vegetable consumption\\
Mummah, S. et al., 2017 \cite{50} &
  N=135 participants, USA &
  Vegethon &
  To examine the effectiveness of a smartphone application to improve the diet's vegetable content among overweight adults endeavoring weight loss.\\
Hull, P. et al., 2017 \cite{51}&
  N=80 participants,USA &
  CHEW &
   
To examine end-user's reviews about the CHEW app regarding usage, usability, perceived limitations, and advantages of the app.\\
Svensson et al., 2016  \cite{52}&
  N= 92 participants, Sweden \mbox{}\protect\newline  &
  FR App \mbox{}\protect\newline  &
  To investigate adolescents' practices using a dietary assessment app, considering circumstances that may impact their dietary intake reporting.\\
Martin, C. K. et al., 2016 \cite{53} \mbox{}\protect\newline  &
  USA &
  Smartloss &
  To demonstrate mHealth platform for weight loss.\\
Zhu, F., et al., 2010 \cite{55} & No participants, USA
   & Dietary Assessment app
   & To accurately measure food and nutrient intake by use of image processing methods.
  \\
Ahmed, M. et al., 2017 \cite{56}  &
  N=18 participants, Canada &
  MyFitness-Pal &
 To evaluate a tablet application's soundness that incorporates detailed nutritional composition data with a calculated food intake/waste activity. \mbox{}\protect\newline \\
Ali, Z. C. et al., 2017 \cite{55-a} &
  N= 54 participants, United Kingdom \mbox{}\protect\newline  &
  Manager App &
  To examine the feasibility of a smartphone application (app.) to calculate nutrient intake as per age-related muscular degeneration\\
Ambrosini, G. L. et al., 2017  \cite{57} &
  N= 50 participants, Australia &
  Easy Diet &
  To examine the application's acceptability and relative validity as an epidemiological diet monitoring tool compared to a conventional dietary assessment method.\\
Bardus, M. et al., 2018 \cite{58} \mbox{}\protect\newline  &
  N= 5,486 participants, Lebanon \mbox{}\protect\newline  &
  WaznApp &
   \mbox{}\protect\newline To investigate the usefulness of a self-directed weight loss intervention and study the outcomes of the intervention. \\
Bennett, Gary G. et al., 2018   \cite{59} &
  N=351 participants,USA &
  Track App &
  To study the effect of the digital weight-loss intervention by embedding it within a community health center system.\\
Brindal, E. et al., 2013 \cite{74}  &
  N= 58 participants, Australia &
  MRP Support App &
  To support individual's partial meal replacement program using a smartphone app.\\
Chen, J. et al., 2019 \cite{61} &
  N= 43 participants, Australia &
  MyFitness-Pal &
  To evaluate how participants in naturalistic environments achieved when reporting their dietary intake and their usability practices.\\
De Cock, N. et al., 2017 \cite{62}\mbox{}\protect\newline  &
  N=889 participants, Belgium &
  Nike+ Running and Fitness Pall etc. &
  To evaluate the association  of commercial  nutrition or fitness apps with a lower value of BMI and healthier intake of snacks and drinking practices in youngsters\\
Everett, E. et al., 2018 \cite{63} &
  N = 55 participants,USA &
  Sweetch &
  To study the acceptability, feasibility, effectiveness, and safety  of the Sweetch app in alliance with an electronic  scale for measuring body weight.\\
Hezarjaribi, N. et al., 2018 \cite{64} &
  N= 30 participants, USA&
  Speech2-Health &
  Examine the technology adaptation rate of the voice-based mobile nutrition monitoring method\\
Ipjian, Michelle  et al., 2017 \cite{65}  \mbox{}\protect\newline  &
  N=30 healthy adults, USA &
  MyFitness-Pal &
  To decrease sodium intake and to study whether a commercial health app is beneficial for supporting dietary change.\\
Jimoh, F. et al., 2018  \cite{66} &
  N=34 participants, United Kingdom &
  FoodWiz2 &
 To study the benefits of a smartphone app with regards to encouraging healthy lifestyle decisions among adolescence.\\
Liu, Y. C. et al., 2019 \cite{67} &
  N= 105 total participants, Taiwan &
  MRP Support App &
   To evaluate the accuracy and time performance of two prototypes for dietary recording utilization\\
Pagoto, S. et al., 2018 \cite{68}&
  Two iterative single arm pilots with participants N=27 and N=16, USA &
  Habit App &
  To automate problem-solving therapy for weight loss using smartphone apps.\\
Paulsen, M. M. et al., 2018 \cite{69} &
  N= 32 hospitalized patients, USA &
  MyFood App &
  To evaluate the app's capability to calculate intake of energy, protein, and liquid with the help of hospitalized patients' food and beverage details.\\
Recio-Rodriguez, et al. J. I., 2019 \cite{70} &
  N=362 individuals, Spain &
  EVIDENT-II App   &
   To compare the calculated values of energy intake, macronutrients, micronutrients, and alcohol consumption using the dietary app with values evaluated using food frequency questionnaire.\\
Shaw, R.et al., 2019 \cite{71}\mbox{}\protect\newline  &
  N= 96 participants, USA &
  My Fitness Pal (Log2Lose) &
  Determine smartphone apps' effectiveness for weight loss programs by giving financial rewards and dietary self-monitoring.

\end{longtable}
\endgroup

\clearpage
\begin{table*}[!ht]
\caption{{Data extraction for answering research question RQ1} }
\label{tw-6aed7cb5ec22}
\def\arraystretch{1}
\ignorespaces 
\centering 
\begin{tabulary}{\linewidth}{p{\dimexpr.4213999999999999\linewidth-2\tabcolsep}p{\dimexpr.5786\linewidth-2\tabcolsep}}
\toprule Attribute & Data capture\\
\midrule 
Difficulty in portion size estimation &
   Resolved when the method to solve the problem is stated or subject can easily key in portion size, Not resolved, Not mentioned\\
Involved too many steps in data entering &
  Subject feedback on the app (Resolved, Not resolved, Not mentioned)\\
Database is not localized &
  Resolved when database region or culture information is specified, Not Resolved, Not mentioned\\
Credibility of database sources &
  Resolved when database sources are listed, Not Resolved, Not mentioned\\
Required to be connected to other electrical devices &
  Resolved when there are no such requirements, Not Resolved, Not Mentioned\\
The user felt the burden to key in foods for every meal &
  Subject feedback on the app (Resolved, Not resolved, Not mentioned)\\
Many information and steps required &
  Subject feedback on the app (Resolved, Not resolved, Not mentioned)\\
Frequent notifications that caused a burden to the user &
  Subject feedback on the app (Resolved, Not resolved, Not mentioned)\\
Demotivating due to the information displayed &
  Subject feedback on the app (Resolved, Not resolved, Not mentioned)\\
App Frequently crashed &
  Resolved when the subject does not face frequent crashes, Not resolved, Not mentioned\\

\bottomrule 
\end{tabulary}\par 
\end{table*}
\begin{table*}[!ht]
\caption{{Data extraction for answering research question RQ2} }
\label{tw-8cd9fe71280b}
\def\arraystretch{1}
\ignorespaces 
\centering 
\begin{tabulary}{\linewidth}{p{\dimexpr.3425000000000001\linewidth-2\tabcolsep}p{\dimexpr.6575\linewidth-2\tabcolsep}}
\toprule Attribute & Data capture\\
\midrule 
Usage reports &
  Feedback on user engagement with the app, Not Mentioned\\
Data confidentiality &
  Resolved when the app specifically addressed the data security issue, Not resolved, Not mentioned\\
Stability of application &
  Resolved, Not Resolved when subject mentioned the bugs and technical issues\\
Offline accessibility &
  Yes when the app does not require internet connection, No)\\
\bottomrule 
\end{tabulary}\par 
\end{table*}
\begin{table*}[!ht]
\caption{{Data extraction for answering research question RQ3} }
\label{tw-2e7650d47b96}
\def\arraystretch{1}
\ignorespaces 
\centering 
\begin{tabulary}{\linewidth}{p{\dimexpr.4493500000000001\linewidth-2\tabcolsep}p{\dimexpr.5506499999999999\linewidth-2\tabcolsep}}
\toprule Attribute & Data capture\\
\midrule 
Design of the app in terms of simplicity, clear and colourful  &
  Keypoints of the detail, Not mentioned\\
Included tutorial page &
  Yes, No, Not Mentioned\\
Include two user dashboards &
  Yes if separate dashboard for patients and more detailed dashboard for dietitians or researchers, No, Not Mentioned\\
User friendly &
  Subject feedback on the mHealth application, Not mentioned\\
Notification to give alert to the user &
  Message details, No, Not mentioned\\
\bottomrule 
\end{tabulary}\par 
\end{table*}
\begin{table*}[!ht]
\caption{{Data extraction for answering research question RQ4} }
\label{tw-e4bf42e0a922}
\def\arraystretch{1}
\ignorespaces 
\centering 
\begin{tabulary}{\linewidth}{p{\dimexpr.2623\linewidth-2\tabcolsep}p{\dimexpr.7377\linewidth-2\tabcolsep}}
\toprule Attribute & Data capture\\
\midrule 
Evaluation of diet quality &
  Method to evaluate diet quality\\
History tracking, and storage of diet records &
  Information regarding diet history, Not mentioned\\
Inclusion of comprehensive database &
  Database name if generic, No (when the database of specific food category or item), Not mentioned\\
Items included in diet record &
  Details of the items in diet record, Not mentioned\\
Methods to estimate portion size &
  Method used such as grams and household measurement and integrated camera, No, Not mentioned\\
Methods to display nutrient components &
  Key points of the method\\
Nutrient component to be included &
  Details of the nutrient components such as calories, protein, fat etc.\\
Validity of database  &
  Name of the database if validated, No, Not mentioned\\
Include to record total water intake &
  No, Yes\\
Inclusion of nutrient information in the database &
  Key details of nutrient information such as calories, fat etc.\\
Provide diet guide that cater to individual needs &
  Key details of method such as notifications, FB messages etc, No, Not mentioned\\
Types of nutrients required to be displayed  &
  Key details of the nutrients displayed in the app, Not mentioned\\
Ways to display summarized energy and nutrient intake &
  Yes and key details, Not mentioned\\
\bottomrule 
\end{tabulary}\par 
\end{table*}

\begingroup
\begin{small}
\begin{table*}[!htbp]
\caption{{(A) Overview of mHealth applications reporting on  general issues (Data extracted for RQ1)} }
\label{table:table3}
\ignorespaces 
\centering 
\begin{tabular}{p{3cm}lp{2.5cm}p{2.5cm}p{2.5cm}p{2.5cm}}


\hline
Application Name&	Year&	Credibility of database sources&	Database is not localized&	Involved too many steps in data entering&	Difficulty in portion size estimation\\ \hline
					
Social POD \cite{16} &
2016&	Not resolved	&Not resolved&	Resolved	&Not mentioned \\ \hline
Happy \cite{18} 
&2017&	Not mentioned	&Not mentioned&	Resolved	&Not mentioned \\ \hline
Lose it \cite{19} 
&2017	&Not resolved&	Not resolved	&Resolved	&Not mentioned \\ \hline
mDPP \cite{20} &
2015	&Not mentioned	&Not mentioned&	Resolved&	Not mentioned \\ \hline
Metabolic Diet app \cite{21} &
2016	&Resolved	&Resolved	&Resolved	&Not resolved \\ \hline
DietCam \cite{22}  
&2012&	Resolved	&Resolved	&Not resolved	&Resolved \\ \hline
My Fitness Pal \cite {23}
&2017	&Not resolved	&Not resolved	&Resolved	&Not resolved \\ \hline
mpFR \cite{25} &
2010	&Resolved	&Not resolved	&Not resolved	&Resolved \\ \hline
FTA \cite{27} &
2014	&Not mentioned	&Not mentioned	&Resolved	&Not resolved \\ \hline
TreC-LifeStyle \cite{28} &
2017&	Not resolved&	Not resolved	&Resolved	&Not resolved \\ \hline
Snack Track School \cite{29} &
2016&	Resolved	&Not resolved	&Resolved	&Not resolved \\ \hline
My Meal Mate \cite {6} &
2013&	Not resolved	&Resolved	&Not resolved&	Not resolved \\ \hline
FRapp \cite{31} &
2015	&Not mentioned&	Not mentioned&	Resolved	&Resolved \\ \hline
iDAT \cite{32} &
2015&	Not resolved&	Not resolved&	Resolved&	Not resolved \\ \hline
MDFR \cite{33} &
2012&	Not mentioned&	Not mentioned	&Resolved	&Resolved \\ \hline
WIZE \cite{73} &
2015&	Not mentioned	&Not mentioned&	Resolved	&Not resolved \\ \hline
ENGAGED \cite{35} & 
2012	&Resolved	&Not resolved	&Resolved	&Not resolved \\ \hline
EVIDENT Mobile app \cite{36}\cite{70} &
2016	&Not resolved	&Not mentioned&	Resolved	&Not resolved \\ \hline
GoCARB \cite{37} &
2016&	Resolved&	Resolved&	Resolved&	Resolved \\ \hline
CalFit Chi \cite{38} &
2016	&Not mentioned&	Not mentioned	&Resolved	&Resolved \\ \hline
APPETITT \cite{43} &
2017	&Not resolved	&Resolved	&Resolved	&Not mentioned \\ \hline
DietApp \cite{44} &
2017&	Not resolved&	Not mentioned	&Not mentioned	&Not mentioned \\ \hline
e-DIA \cite{47} &
2016	&Resolved	&Resolved	&Not resolved&	Not mentioned \\ \hline
Vegethon \cite{49} &
2017	&Not mentioned&	Resolved	&Resolved	&Resolved \\ \hline
Dietary Intake Assessment \cite{52} &
2016	&Resolved	&Resolved&	Not resolved	&Not resolved \\ \hline
MRP Support App \cite{74} &
2013	&Not mentioned	&Not mentioned&	Resolved&	Resolved \\ \hline
Nutricam \cite{75} &
2011&	Not mentioned&	Not mentioned	&Resolved	&Resolved \\ \hline
Easy Diet \cite{57} &
2017&	Resolved	&Not resolved&	Resolved	&Not mentioned \\ \hline
Lark \cite{58} &
	2018&	Resolved&	Resolved&	Resolved	&Not mentioned \\ \hline
Sweetch \cite{63} &
2018&	Not mentioned	&Resolved	&Resolved &	Not mentioned \\ \hline
Speech2Health \cite{64} &
2018&	Resolved&	Not resolved&	Resolved	&Resolved \\ \hline
FoodWiz2 \cite{66} &
2018	&Resolved&	Not resolved&	Resolved	&Resolved \\ \hline
Habit App \cite{68} &
2018	&Resolved&	Not resolved&	Resolved	&	Not mentioned \\ \hline
MyFood App \cite{69} &
2018&	Not resolved	&Not resolved&	Resolved& 	Resolved \\ \hline
MyFitness Pal (Log2Lose) \cite{71} &
2019&	Not resolved 	&Not resolved	&Resolved 	&Not resolved \\ 
\tblbottomrule 
\end{tabular}\par 
\end{table*}
\end{small}
\endgroup
\textbf{\space }
\clearpage
\begin{landscape}

\makeatletter\@twocolumnfalse\makeatother
\begingroup
\makeatletter\if@twocolumn\@ifundefined{theposttbl}{\gdef\TwoColDocument{true}\onecolumn\onecolumn}{}\fi\makeatother \setlength\LTcapwidth{\textheight}
\begin{longtable}{p{2.0cm}p{1.0cm}p{2.0cm}p{2.0cm}p{2.0cm}p{1.8cm}p{2.3cm}p{2.1cm}p{2.4cm}}
\caption{{\textbf{(B) Overview of mHealth applications reporting on general issues (Data extracted for RQ1).}} }
\label{table:table4}
\def\arraystretch{1}\\\endfirsthead \hline \noalign{\vskip3pt} \noalign{\textit{Table \thetable\ continued}} \noalign{\vskip3pt} \endhead \noalign{\vskip3pt} \noalign{\textit{\hfill Continued on next page}} \noalign{\vskip3pt} \endfoot \endlastfoot

\\
\hline
Application Name	&Year	&App frequently crashed	&Demotivated due to information displayed&	Frequent notifications that caused burden to user&	Many information and steps required&	The user felt the burden to key in foods for every meal&	Required to be connected to other electrical devices&	Difficulty in portion size estimation \\ \hline

Social POD \cite{16} &
2016&	Resolved	&Resolved	&Resolved	&Resolved	&Resolved	&Resolved	&Not mentioned \\ \hline
Happy \cite{18} &
2017	&Not resolved	&Resolved	&Resolved	&Resolved	&Resolved&	Resolved&	Not mentioned \\ \hline
Lose it \cite{19} &
2017&	Not resolved	&Not resolved	&Not resolved	&Resolved	&Resolved	&Resolved	&Not mentioned \\ \hline
mDPP \cite{20} &
2015&	Resolved	&Resolved	&Resolved	&Resolved	&Not resolved	&Resolved	&Not mentioned \\ \hline
Metabolic diet app \cite{21} 
&2016	&Resolved	&Resolved	&Not mentioned	&Resolved	&Resolved	&Resolved	&Not resolved \\ \hline
DietCam \cite{22}&
2012&	Resolved&	Resolved	&Resolved	&Not resolved	&Not resolved	&Resolved	&Resolved  \\ \hline
My Fitness Pal \cite {23} &
2017&	Resolved&	Resolved&	Resolved	&Resolved	&Resolved&	Resolved&	Not resolved  \\ \hline
mpFR \cite{25} &
2010&	Resolved	&Resolved&	Not mentioned&	Not resolved&	Not resolved&	Resolved&	Resolved  \\ \hline
FTA \cite{27} &
2014	&Resolved&	Not resolved&	Not mentioned	&Resolved&	Not resolved&	Resolved&	Not resolved \\ \hline
TreC-LifeStyle \cite{28} &
2017&	Resolved	&Resolved&	Resolved&	Resolved	&Resolved	&Resolved&	Not resolved \\ \hline
Snack Track School \cite{29} &
2016	&Resolved	&Resolved	&Resolved	&Resolved	&Resolved	&Resolved	&Not resolved  \\ \hline
My Meal Mate \cite{6}\cite{27} &
2013	&Resolved&	Resolved&	Not mentioned&	Not resolved	&Not resolved	&Resolved&	Not resolved  \\ \hline
FRapp \cite{31} &
2015&	Resolved	&Resolved&	Not mentioned&	Not resolved	&Not resolved&	Resolved	&Resolved \\ \hline
iDAT \cite{32} &
2015&	Resolved	&Resolved&	Not mentioned	&Resolved	&Not resolved&	Resolved	&Not resolved \\ \hline
MDFR \cite{33} &
2012&	Resolved	&Resolved	&Not resolved	&Resolved	&Resolved	&Resolved	&Resolved \\ \hline
WIZE \cite{73} &
2015&	Resolved	&Not resolved	&Not mentioned	&Resolved	&Resolved &	Resolved	&Not resolved \\ \hline
ENGAGED \cite{35} &
2012&	Resolved&	Resolved&	Resolved&	Resolved&	Resolved&	Resolved	&Not resolved \\ \hline
EVIDENT Mobile app \cite{36}\cite{70} &
2016	&Resolved&	Resolved	&Not mentioned	&Resolved&	Resolved&	Resolved	&Not resolved \\ \hline
GoCARB \cite{37} &
2016	&Resolved&	Resolved	&Not mentioned	&Resolved&	Resolved&	Resolved	&Resolved \\ \hline
Calfit Chi \cite{38} &
2016&	Resolved&	Resolved&	Resolved&	Not resolved&	Resolved	&Resolved&	Not resolved \\ \hline
								
APPETITT \cite{43} &
2017&	Not mentioned&	Resolved	&Resolved&	Resolved&	Resolved	&Resolved	&Not mentioned \\ \hline
DietApp \cite{44} &
2017&	Not mentioned	&Resolved&	Not mentioned&	Not mentioned	&Resolved&	Resolved&	Not mentioned \\ \hline
e-DIA \cite{47} &
2016	&Not mentioned&	Not mentioned	&Not mentioned	&Not resolved	&Not mentioned&	Resolved&	Not resolved \\ \hline
Vegethon \cite{49} &
2017&	Not mentioned&	Resolved	&Resolved&	Resolved	&Resolved	&Resolved	&Resolved \\ \hline
Dietary Intake Assesment \cite{52}&
2016	&Resolved	&Resolved& 	Not mentioned&	Not resolved&	Not resolved&	Resolved	&Not resolved \\ \hline
MRP Support App \cite{74} &
2013&	Not mentioned&	Resolved&	Resolved	&Resolved	&Resolved&	Resolved&	Resolved \\ \hline
Nutricam \cite{75} &
2011&	Not mentioned	&Resolved	&Not mentioned	&Resolved	&Resolved	&Resolved&	Resolved \\ \hline
Easy Diet \cite{57} &
2017&	Resolved&	Resolved&	Resolved&	Resolved&	Resolved&	Resolved&	Not mentioned \\ \hline
Lark \cite{58} &
2018&	Resolved &	Resolved&	Resolved&	Resolved&	Not resolved&	Resolved&	Not mentioned \\ \hline
Sweetch \cite{63} & 
2018&	Resolved&	Resolved&	Resolved&	Resolved&	Not resolved&	Resolved&	Not mentioned \\ \hline
Speech2Health \cite{64} &
2018&	Resolved&	Not mentioned&	Not mentioned&	Resolved&	Resolved&	Resolved&	Resolved \\ \hline
FoodWiz2 \cite{66} &
2018&	Resolved&	Resolved&	Resolved&	Resolved&	Not resolved&	Not resolved&	Resolved \\ \hline
Habit App \cite{68} &
	2018&	Resolved&	Resolved&	Resolved&	Resolved&	Not resolved&	Resolved&	Not resolved \\ \hline
MyFood App \cite{69} &
2018&	Resolved&	Resolved&	Resolved&	Resolved&	Resolved&	Resolved&	Resolved \\ \hline
My Fitness Pal (Log2Lose) \cite{71} &
2019&	Resolved&	Resolved &	Resolved&	Resolved&	Resolved&	Not resolved&	Not resolved \\ 
\tblbottomrule 
\end{longtable}
\endgroup
\clearpage
\makeatletter\@ifundefined{TwoColDocument}{}{\twocolumn}\makeatother 
\end{landscape}
 \begingroup
\begin{small}
\begin{landscape}
\makeatletter\@twocolumnfalse\makeatother
\begingroup
\makeatletter\if@twocolumn\@ifundefined{theposttbl}{\gdef\TwoColDocument{true}\onecolumn\onecolumn}{}\fi\makeatother \setlength\LTcapwidth{\textheight}
\begin{longtable}{p{3cm}lp{1.5cm}p{4cm}p{4cm}p{6cm}}
\caption{{\textbf{Overview of mHealth applications reporting on the Backend issues (Data extracted for RQ2).}} }
\label{table:table5}
\def\arraystretch{1}\\\endfirsthead \hline \noalign{\vskip3pt} \noalign{\textit{Table \thetable\ continued}} \noalign{\vskip3pt} \hline \endhead \hline \noalign{\vskip3pt} \noalign{\textit{\hfill Continued on next page}} \noalign{\vskip3pt} \endfoot \endlastfoot 

\hline
Application Name&	Year&	Offline accessibility&	Stability of application&	Data Confidentiality&	Usage Reports \\ \hline
					
Social POD \cite{16} &
2016&	No&	Resolved&	Not mentioned&	User engagement declined over time \\ \hline
Happy \cite{18} & 2017&	No&	Not resolved(Bugs and technical issues)&	Not mentioned	&Engagement with the app was low \\ \hline
Lose it \cite{19} &
2017&	No&	Not resolved (App crash and bugs)&	Not mentioned&	Not mentioned \\ \hline
mDPP \cite{20} &
2015&	No&	Not resolved (App glitches)&	Not mentioned&	Declined Adherence to app overtime \\ \hline
Metabolic Diet app \cite{21} &
2016 &	No &	Resolved&	Not mentioned&	Not mentioned \\ \hline
Diet Cam \cite{22} &
2012 &	No&	Resolved&	Not mentioned&	Not mentioned \\ \hline
My Fitness Pal \cite {23} &
2017 &	Yes&	Resolved &	Not mentioned&	Not mentioned \\ \hline
mpFR \cite{25} &
2010&	No&	Resolved&	Not mentioned&	Not mentioned \\ \hline
FTA \cite{27} &
2014&	No&	Not resolved&	Resolved (Secure server)& 	Not mentioned \\ \hline
TreC-LifeStyle \cite{28} &
2017&	No&	Resolved&	Resolved  &		Good usability with high adherence. \\ \hline
Snack Track School app \cite{29} &
2016&	No&	Resolved&	Not mentioned&Good usability with the weekly report to measure user engagement.    \\ \hline
My Meal Mate
\cite {6} &2013&	No&	Not resolved &	Not mentioned&	Apps adherence declined over time.
\\ \hline
FRapp \cite{31} &
2015&	No&	Resolved&	Not mentioned&	Not mentioned \\ \hline
iDAT \cite{32} &
2015&	No&	Resolved&	Resolved (Personal information \& data were kept confidential)&	Usability is good and app usage is tracked continuously  \\ \hline
MDFR \cite{33} &
2012&	No&	Resolved&	Not mentioned&	Not mentioned \\ \hline
WIZE \cite{73} &
2015&	No&	Not resolved (User interface and design issues) &	Not mentioned&	Not mentioned \\ \hline
 ENGAGED \cite{35} &
2012&	No&	Resolved&	Resolved &	Good usability and usage access reports based on user reporting.\\ \hline
EVIDENT Mobile phone app \cite{36}\cite{70} &
2016 &	No&	Resolved&	Not mentioned&	The adherence score is based on user records. \\ \hline
GoCARB \cite{37} &
2016&	No&	Not resolved&	Resolved (Images are transmitted to a dedicated secure server)&	Not mentioned \\ \hline
CalFit Chi \cite{38} &
2016&	No&	Resolved&	Resolved (Data are encrypted and stored locally in the memory on the phone)&	Not mentioned \\ \hline
APPETITT \cite{43} &
2017&	Yes&	Resolved&	Not mentioned&	Not mentioned  \\ \hline
DietApp \cite{44} &
2017&	No&	Resolved&	Not mentioned&	Not mentioned \\ \hline
e-DIA \cite{47} &
2016&	No&	Resolved&	Not mentioned&	Not mentioned \\ \hline
Vegethon \cite{49} &
2017&	No&	Resolved&	Not mentioned&	Not mentioned \\ \hline
Dietary Intake Assessment \cite{52} &
2016&	No&	Resolved&	Not mentioned	&
The researcher can view user profile data, food recordings, and user responses to questions. \\ \hline
MRP Support App \cite{74} &
2013&	No&	Resolved&	Not mentioned	&Not mentioned \\ \hline
Nutricam \cite{75} &
2011&	No&	Resolved&	Not mentioned&	Not mentioned\\ \hline
Easy Diet \cite{57} &
 2017&	No&	Not resolved&	Not mentioned&	Continuous feedback and high engagement  \\ \hline
Lark \cite{58} &
2018&	No&	Resolved&	Secure Server&	Continuous feedback on behavior out comes using Artificial Intelligence helps to engage users  \\ \hline
Sweetch  \cite{63} &
2018&	No&	 Resolved&	Not mentioned&	Personalized notifications engages users of the app. \\ \hline
Speech2Health \cite{64} &
2018&	No&	Not resolved&	Not mentioned&	Not mentioned \\ \hline
FoodWiz2 \cite{66} &
2018	&No&	Resolved&	Not mentioned&	 Personalized feedback and help increase motivation and user engagement. \\ \hline
Habit App \cite{68} &
2018&	No	 &Resolved	&Not mentioned	&Adherence to the app was high as it helps users to lose weight by resolving the problems  \\ \hline
MyFood App \cite{69} &
2018&	No&	Resolved&	Resolved (Encrypted Data Transfer)&	High adherence and the majority of users can learn new information.  \\ \hline
My Fitness Pal (Log2Lose) \cite{71} &
2019&	Yes&	Resolved&	Not mentioned&	Usability is high as apps offers financial incentives for logging foods and achieving goals
\\ 
\tblbottomrule 
\end{longtable}
\endgroup
\makeatletter\@ifundefined{TwoColDocument}{}{\twocolumn}\makeatother 
\end{landscape}
\end{small}

\clearpage
\makeatletter\if@twocolumn\@ifundefined{theposttbl}{\gdef\TwoColDocument{true}\onecolumn\onecolumn}{}\fi\makeatother \setlength\LTcapwidth{\textwidth}
\begin{small}

\begin{longtable}{p{3cm}lp{3cm}p{2.5cm}p{2.5cm}p{3cm}}

\caption{{\textbf{(A)Overview of mHealth apps reporting on user interface requirement (Data extracted for RQ3). }} }

\label{table:table6}
\def\arraystretch{1}\\\endfirsthead \hline \noalign{\vskip3pt} \noalign{\textit{Table \thetable\ continued}} \noalign{\vskip3pt} \hline \endhead \hline \noalign{\vskip3pt} \noalign{\textit{\hfill Continued on next page}} \noalign{\vskip3pt} \endfoot \endlastfoot

\hline		
Application Name&	Year&	Design of the app&	Include tutorial page&	Include two user dashboards	&User friendly \\ \hline
Social POD \cite{16} &
2016 &	Not mentioned &	Not mentioned &	Not mentioned &	Not mentioned \\ \hline
Happy \cite{18} & 
2017 &	Simple, font size is medium &	Not mentioned &	Not mentioned&	Easy to use app \\ \hline
Lose it \cite{19} &
2017 &	Font size is okay, clear images and relevant icon &	Not mentioned &	Not mentioned &	Less step to key in data \\ \hline
mDPP \cite{20} &
2015 &	Not mentioned	 &Not mentioned	 &Not mentioned &	App glitches,  \\ \hline
Metabolic Diet app \cite{21} &
2016 &	Simple, relevant icon, nice color, clear font  &	Specify preferred IEM diet app, option for users to enter weight  &	Personalized dashboard according to preferred IEM diet&	User friendly interface, the use of lay language and distinctive icons \\ \hline
Diet Cam \cite{22} &
2012 &	Simple design,	 &Not mentioned	&Calorie, camera, calendar and album tab	&Easy to operate with few tabs \\ \hline
My Fitness Pal \cite {23} &
2017	 &Simple and Interactive Design &	Yes &	Mobile and Web Platform	 &Easy to use app \\ \hline
mpFR \cite{25} &
2010 &	Very simple design &	Not mentioned	 &Not mentioned &	Less steps \\ \hline
FTA \cite{27} &
2014 &	Simple design, large icon & 	Not mentioned & 	Food habit registration system &	Cumbersome user interfaces \\ \hline
TreC-LifeStyle \cite{28} &
2017 &	Simple, icons and clear font	 &Not mentioned	 &Web and mobile platform  &	Easy to use\\ \hline
Snack Track School app \cite{29} &
2016 &	Not mentioned &	Not mentioned &	Not mentioned &	Not mentioned\\ \hline
My Meal mate \cite{6}\cite{30} &
2013 &	Simple, less icon, large font &	Not mentioned&	Mobile and Web Platform&	Encountered bug but easy to use \\ \hline
FRapp \cite{31} &
2015 &	Very simple, not attractive, large font, less icon \& images&	Not mentioned&	Mobile app&	Not mentioned \\ \hline
iDAT \cite{32} &
2015&	Simple design, nice icon, font and image clear&	Not mentioned	&Mobile App	&Easy to use and interactive \\ \hline
MDFR \cite{33} &
2012 &	Simple interface&	Not mentioned&	Mobile app&	Not mentioned \\ \hline
WIZE \cite{73} &
2015 &	Simple and basic, large icon  &	Not mentioned &	Mobile app	&Cumbersome to use\\ \hline
 ENGAGED \cite{35} &
2012 &	Simple interface, less attractive, clear font &	Not mentioned	&Mobile app and coaching application	&Simple interface \\ \hline
EVIDENT Mobile phone apps \cite{36,70} &
2016 &	Not mentioned &	Not mentioned &	Mobile app	&Easy to use interface for logging food \\ \hline
GoCARB \cite{37} &
2016 &	Simple, big icons, small font&	Not mentioned	&Mobile app	&Graphical user interface, Easy to use \\ \hline
CalFit Chi \cite{38} &
2016 & 	Not mentioned &	Not mentioned &	Mobile app	 &Not mentioned \\ \hline
APPETITT \cite{43} &
2017 &	Simple user interface without menu function, distinctive color contrasts, and large letters.&	No	&Not mentioned&	Yes \\ \hline
DietApp \cite{44} & &
	Simple design&	No&	Not mentioned&	Yes. 84\% of users
were of the opinion that the app was easy to use \\ \hline
e-DIA \cite{47} &
2016 &	Simple \& clean	&No	&No	&Yes \\ \hline
Vegethon \cite{49} & 2018 &
	Uses icon, graphic, simple and colorful design&	Yes&	Not mentioned&	High rate of engagement. \\ \hline
Dietary Intake Assessment \cite{52} &
2016&	Not mentioned&	Not mentioned&	Mobile and Web Platform& 	Most Users found it was easy to record dietary intake see the results, but there were also views that the recording of dietary intake was burdensome. \\ \hline
MRP Support App \cite{74} &
2013&	Simple interface, more icon \& less word. &	Yes	&Not mentioned&	Yes\\ \hline
Nutricam \cite{75} &
2011&	Simple interface&	No&	No	&Yes \\ \hline
Easy Diet \cite{57} &
2017&	Simple and interactive interface&	Not mentioned&	Mobile app&	Easy to use app\\ \hline
Lark \cite{58} &
2018&	Simple Interface&	Yes	&Mobile App	&Easy to use app \\ \hline
Sweetch \cite{63} &
2018&	Simple And Interactive Interface&	Not mentioned&	No&	83 percent are of opinion that app was easy to use \\ \hline
Speech2Health \cite{64} &
2018&	Simple And Interactive Interface&	Not mentioned&	 No	&High rate of  user friendliness as user adds  meal using speech to text conversation \\ \hline
FoodWiz2 \cite{66} &
2018&	Simple Interface&	Not mentioned&	Mobile and web platform&	Easy to use as users are able to add meal using barcode scanning  \\ \hline
Habit App \cite{68} &
2018&	Simple and interactive interface&	Yes&	Mobile and Web Platform&	The high rate of engagement  as it introduced the problem-solving process for weight loss \\ \hline
MyFood App \cite{69} &
2018&	Simple and interactive interface&	Not mentioned&	Mobile and Web Platform&	Ninety percent of participant reported that the app was easy to use.  \\ \hline
My Fitness Pal
(Log2Lose) \cite{71} &
2019&	Simple and interactive design& 	Yes&	Mobile and Web Platform&	High Rate of engagement as it offers financial incentives for logging food  \\

\hline
\end{longtable}
\end{small}
\endgroup
\clearpage
\makeatletter\@ifundefined{TwoColDocument}{}{\twocolumn}\makeatother 

\begin{landscape}
\makeatletter\@twocolumnfalse\makeatother
\begingroup
\makeatletter\if@twocolumn\@ifundefined{theposttbl}{\gdef\TwoColDocument{true}\onecolumn\onecolumn}{}\fi\makeatother \setlength\LTcapwidth{\textheight}
\begin{small}
\begin{longtable}{p{3cm}lp{4.3cm}p{3.3cm}p{3.5cm}p{5cm}}
\caption{{\textbf{(B) Overview of mHealth apps reporting on the user interface requirement (Data extracted for RQ3).}} }
\label{table:table7}
\def\arraystretch{1}\\\endfirsthead \hline \noalign{\vskip1pt} \noalign{\textit{Table \thetable\ continued}} \noalign{\vskip1pt} \hline \endhead \hline \noalign{\vskip1pt} \noalign{\textit{\hfill Continued on next page}} \noalign{\vskip1pt} \endfoot \endlastfoot

\hline
Application Name&	Year&	Design of app in terms of simplicity, clear and colorful&	Items to include in the user profile&	Notification to give alert to user&	Things to be considered as user friendly  \\ \hline
Social POD \cite{16} &
2016&	Not mentioned&	User Goals&	Message send to user who did not enter data for previous 48H	& Saving a meal as a template to be loaded the next time when an identical meal is consumed; User could take a picture of their meal as a memory aid if the consumed foods could not be entered until later \\ \hline
Happy \cite{18} &
2017&	Simple, soft color and clear images &	Not mentioned	& Tailored messages sent to user everyday &	It is easy to use and learn to operate.
Messages sent easy to understand, and meaningful information \\ \hline
Lose it \cite{19} &
2017&	Clear images, simple design, easy to use &	Dietary goals, &	FB messages generated (software) &	Easy to use and less step for data entry \\ \hline
mDPP \cite {20} &
2015 &	Not mentioned	 &Not mentioned &	Automated text message was generated as reminder &	App glitches reflect adherence \\ \hline
Metabolic Diet app \cite{21} &
2016 &	Simple design with soft color and clear icon/words &	Body weight, preferred IEM apps	& Not mentioned	 &User friendly interface, the use of lay language and distinctive icons \\ \hline
Diet cam \cite{22}  &
2012 &	Simple, not attractive	&Not mentioned&	No&	Easy to operate with few tabs \\ \hline
My Fitness Pal [23] [56] [61] [65] &
2017 &	Not mentioned &	Allow user to set weight and nutrition goals	&Feedback on the number of calories and nutrient needed to reach goal&	User friendly interface \\ \hline
mpFR \cite {25} &
2010&	Very simple design (Interaction design)	& Not mentioned&	No	& Easy to use \\ \hline
FTA \cite {27} &
2014&	Clear images, nice color, simple design&	Management of personal goals&	No&Cumbersome user interface	 \\ \hline
TreC-Lifestyle \cite {28} &
2017&	Simple with soft colors clear icons and words&	Configuration of an app based on children's age and weight&	Feedback and virtual coaching function&High adherence of usage	 \\ \hline
Snack Track School app \cite{29} &
2016&	Not mentioned&	Goal Setting&	No&	Not mentioned \\ \hline
My Meal Mate \cite {6}\cite{27}&
2013&	Design and color not attractive&	Goal setting&	Feedback via weekly message	&Easy to use  \\ \hline
FRapp \cite{31} &
2015&	Simple design, relevant color, clear icon&	Not mentioned&	No&	Camera function with easy interface \\ \hline
iDAT \cite{32} &
2015&	Color too soft, icon and words are clear&	Weight loss goal&	Reminder&	Easy to use, interactive \\ \hline
MDFR \cite{33} &
2012&	Very simple, less function&	Not mentioned&	Text message as reminder&	Easy to use \\ \hline
WIZE \cite{73} &
2015&	Simple and basic, colorful &	Goal setting( dietary)&	No	& Personalized dashboard to individual not exist, prompt and reminders, to add pictures and colors. User interface  and design issues \\ \hline
 ENGAGED \cite{35} &
2012&	Black background not suitable, simple interface	&Calorie goals&	Real time feedback and coaching&	Simple interface \\ \hline
EVIDENT Mobile phone apps \cite{36}\cite{70} &
2016&	Not mentioned&	Not mentioned&	No	& Easy to use interface \\ \hline
GoCARB \cite{37} &
2016&	Basic color and simple design&	Not mentioned&	No&	Useful and easy to use \\ \hline
CalFit Chi \cite{38} &
2016&	Not mentioned&	Not mentioned&	Reminder text messages&	- \\ \hline
APPETITT \cite{43} &
2017&	A stable user interface without menu function, distinctive color contrasts and large letters, consistent touch functions and large touch surfaces&	No feature for personalized profile&	APPETITT had a function that notified the user
with a short signal at mealtime&
	Minimal entering of data, everything is on the same screen. Very simple and user friendly\\ \hline
DietApp \cite{44} & 2016&
	Simple and clear design. 43\% users found it attractive. &	Stores name, password, email, weight, height, gender
and any illnesses that the user may suffer from. 	&No	&Covers the most common foods; simple home page that can access all feature just from there; ease of installation and use can be used
by all type of public (experts and non-expert) \\ \hline
e-DIA \cite{47} &
2016 &	Simple \& clean	& Not mentioned&	a text message reminder sent to participant  to record food intake on a dedicated days in a week&	Has a complete dataset for varieties of food. They also can put custom food \\ \hline
Vegethon \cite{49}  & &
	Uses icon, graphic, simple and colorful design	&Not mentioned	&Push notifications: notifications to prompt self-monitoring of vegetable consumption &	Provide tutorial; simple graphic and attractive design; targeted apps (focus on veggie intake only)  \\ \hline
Dietary Intake Assessment \cite{52}  &
2016 &	Not clearly mention	&Name, date of birth, gender, weight, height, email address, and phone number.&	Push notification reminders to record meals at a chosen time interval &	Complete dataset with varieties of foods; 
connected to the mobile phone camera and the user could take a picture of their meal as a memory aid if the consumed foods could not be entered until later;  save a meal as a template to be loaded the next time the same meal was consumed \\ \hline

MRP Support App \cite{74}  &
2013 &	Simple, more icon clean but less attractive	& No personalized user profile&	Push notifications to remind user to record meal in 3 mealtimes&	Clean UI, simple icon, not many step require to log  meal and weight \\ \hline
Nutricam \cite{75} &
2011 &	Simple and clear&	No personalized user profile&	No	&Few step to record food. Very simple app \\ \hline
Easy Diet \cite{57}   &
2017 &	Clear image, simple design, easy to use	&Not mentioned	& Disabled for this study &	Easy to use and less step for data entry. Automatically detects food and beverages \\ \hline
Lark \cite{58} & 

2018 &	Simple And Clear&	 Goals Settings &	Sends notification  to user to review activity throughout the day	& Interactive Conversations \\ \hline
Sweetch \cite{63}  &
2018 &	Simple and clear according to majority of users	& Goal settings &	Push notifications based on habits &	Well Integrated, Consistence, Easy To Learn And Use \\ \hline
Speech2Health \cite{64}  &
2018&	Simple and clear &	Not mentioned	&Feedback on calories and nutrients to reach goal	& Record food  without manual logging by using  speech to text conversation \\ \hline
FoodWiz2 \cite{66} &
2018&	Simple and clear &	Goal settings&	Personalized feedback text messages for motivation and help &	Allow users to record home prepared recipes and food products through barcode scanning \\ \hline
Habit App \cite{68}  &
2018	 &
Simple and clear&	User profile, &Goal settings&	Notifications on problem solving process and Automatic Problem solving process \\ \hline
MyFood App \cite{69} &
2018 &	Interactive design and clear icons with soft colors &	User profile	& Not mentioned	 & Allow users to add food and determine its nutrient content by taking picture  \\ \hline
My Fitness Pal (Log2Lose) \cite{71}  &
2019 &	Not mentioned &	Allow user to set weight and nutrition goals	& User receives messages about financial incentives every week  &	User friendly interface \\ 

\hline
\end{longtable}
\end{small}
\endgroup
\makeatletter\@ifundefined{TwoColDocument}{}{\twocolumn}\makeatother 
\end{landscape}
\clearpage
\begin{landscape}
\makeatletter\@twocolumnfalse\makeatother
\begingroup
\makeatletter\if@twocolumn\@ifundefined{theposttbl}{\gdef\TwoColDocument{true}\onecolumn\onecolumn}{}\fi\makeatother \setlength\LTcapwidth{\textheight}
\begin{small}
\centering
\begin{longtable}{p{1.5cm}lp{2.1cm}p{2.1cm}p{2.4cm}p{2.1cm}p{2.0cm}p{2.4cm}p{2.0cm}p{1.5cm}}
\caption{{\textbf{(A) Overview of mHealth apps reporting on the implementation of the dietary components (Data extracted for RQ4).}} }
\label{table:table8}
\def\arraystretch{1}\\\endfirsthead \hline \noalign{\vskip3pt} \noalign{\textit{Table \thetable\ continued}} \noalign{\vskip3pt} \hline \endhead \hline \noalign{\vskip3pt} \noalign{\textit{\hfill Continued on next page}} \noalign{\vskip3pt} \endfoot \endlastfoot



\hline
Application Name&	Year&	Evaluation of diet quality&	History tracking and storage of diet records&	Inclusion of comprehensive database&	Items included in diet record&	Methods to display nutrient components&	Methods to estimate portion size&	Nutrient component to be included&Validity of database \\ \hline
Social POD \cite{16} &
2016&	To track diet and reach target goal(calorie consumed)&	Able to review history&	No&	Not mentioned&	Calories per meal&	Not mentioned&	Calories only&	Not validated  \\ \hline
Happy \cite{18} &
2017&	The intake of fruit and vegetables&	Trends history&	No	Fruit and vegetables&	Not mentioned&	Not mentioned&	Not mentioned&	No & Not mentioned  \\ \hline
Lose it \cite{19} &
2017&	Achieve dietary goals, calories, fat, sugar( calorie allowance and balances)&	Select date to view history&	Not mentioned&	Not mentioned&	Balance macronutrient&	Manual estimation with household measurement	&Calories, fat and sugar &	Not mentioned \\ \hline
mDPP \cite {20} &
2015&	The intake of calories,SF, sugar( dietary goals)&	Calorie diary&	No	&Not mentioned&	Diary pattern&	Not mentioned&	Calories, saturated fat, sugar&	Not mentioned \\ \hline
Metabolic Diet app \cite {21} &
2016&	Specific nutrient goals for daily goals	&Daily food diary and review food log&	GMDI Metabolic Pro  database with 100,000 food&	Not mentioned&	A pie chart&	Not mentioned	&Nutrient information is based on Metabolic Pro (Energy, CHO, fat, protein, amino acid)&	USDA food database \\ \hline
Diet cam \cite{22} &
2012&	Calorie intake&	Calendar menu view history by date&	Global database and small personal database&	Not mentioned&	Food images with calorie&	Volume estimation from the integrated camera&	Calories&	USDA food database \\ \hline
My Fitness Pal \cite{23}&
2017&	Breakdown of daily calories and nutrient based on goal&	Select date to review history&	Not mentioned&	Not mentioned&	Breakdown of daily calories and nutrient& 	Household measurement&	Calories and nutrients&	Not mentioned \\ \hline
mpFR \cite{25} &
2010&	Nutrient content of food consumed&	Not mentioned&	Food and nutrient database&	Not mentioned&	Labeled image and nutrient analysis&	Integrated camera&	Nutrients&	FNDDS \\ \hline
FTA \cite{27} &
2014&	Personal goals&	Not mentioned&	No&	Not mentioned&	Food diary&	Not mentioned&	Not mentioned&	Not mentioned \\ \hline
TreC-LifeStyle \cite{28} &
2017&	Color bars to achieve balance nutrients and meal&	Select date to view history record&	No&	Not mentioned&	Progress bar of meals and nutrient balance	&Grams&	Calories, carbs, protein, fat	&No \\ \hline
Snack Track School app \cite{29} &
2016&	Points for nutritional value of snack&	Monitor in weekly report&	A large snack database&	Snack&	Nutrition value of snack&	Common standard measure&	Calories&	Based on International Trade Name Database \\ \hline
My Meal Mate \cite{6} &
2013&	Remaining calories (Daily summary) & 	Select date to view history record	& UK-specific branded food database &	Not mentioned	&Breakdown of calories per meal in table&	Common standard measure&	Calorie&	Database provided by Weight Loss resources, a commercial company \\ \hline
FRapp \cite{31} &
2015&	Food record with image	&Tab to review history record&	No &Not mentioned&	Breakdown of daily meal with images&	Integrated camera&	No	&No \\ \hline
iDAT \cite{32} &
2015&	Balanced calorie to consumed&	Tab to review history record&	A database with locally available food&	Not mentioned&	Breakdown of calories per meal, calories consumed progress bar&	Not mentioned&	Calories&	Not mentioned  \\ \hline
mpFR \cite{25} &
2012&	Food record with dietary feedback&	Not mentioned&	No	&Not mentioned	&Food images before and after eating&	Camera	&No&	No \\ \hline
WIZE \cite{73} &
2015&	Goals tracker weekly review	&Goal week review	&No&	Not mentioned&	Weekly review goals in list&	No&	Not mentioned&	No \\ \hline
 ENGAGED \cite{35} &
2012&	Showing remaining calories and fat&	Not mentioned&	Calorie king database&	Not mentioned&	Goal thermometer display calorie and fat consumed&	Not mentioned&	Calories and fat&	Comprehensive nutritional source containing over 50, 000 food entries \\ \hline
EVIDENT Mobile phone apps \cite{36}\cite{70} &
2016&	Personalized recommendation	&Not mentioned&	Not mentioned&	Not mentioned&	Not mentioned&	Not mentioned&	Carb, protein, fats, fiber&	Not mentioned \\ \hline
GoCARB \cite{37} &
2016&	Images and CHO estimation&	Select by date&	USDA nutrient database&	Not mentioned&	Food type, volume and cho estimation in table&	Integrated camera, recognition, volume estimation&	CHO&	USDA \\ \hline
CalFit Chi \cite{38} &
2016&	Not mentioned&	Not mentioned&	Not mentioned&	Not mentioned&	Not mentioned&	Video camera, record contents of meal, coded the portion size by dietitian&	Not mentioned&	Not mentioned \\ \hline
APPETITT \cite{43} &
2017&	Only a graphic chart for displaying goal achievement.&	No function to see what food and drink reported earlier.&
	No food database info &	Not mentioned&	
Graphic chart with the capacity of 4 meals and 20 units of liquid keeps track of food and water intake.	&No portion size info. User only tell whether they eat or not.&	No macro/micronutrient included. Only water and simple meal&	No food database incorporated  \\ \hline
DietApp \cite{44} &
2017&	
Shows total energy balance and  alert the user &	Not stated &	Not mentioned &	Only calories is shown&	Food History is not displayed, and the app alerts the user when taking certain food affects the calorie goal. &	
User can choose the portion size of food 	&Calories shown based on RDI &	Not mentioned \\ \hline
e-DIA \cite{47}  &
2016&	Food displayed in grams, and only the researcher can view other nutrients &24h Food record  only stored  in the phone and other Data is stored in the cloud  &	Using Australian food composition database &	Only food item with gram	& Simple table&	User estimate portion sizes using metric cups and spoons with manual booklet before entering gram of food	&Gram only &	Database used was AUSNUT 2007. \\ \hline
Vegethon \cite{49}  &
2017  &	Only focus on veggie intake	& Store veggie intake record &	Not clearly mention what database used&	Only servings sizes of veggies	S&how plate graphic icon of veggies serving sizes&	User estimate using ‘1 serving’ standard of veggies&	Not mentioned&	Not mentioned \\ \hline
Dietary Intake Assessment \cite{52}  &
2016 &The user can see details of TEE,  EI, food in grams, dietary fiber, calcium, iron, vitamin C, vitamin D, and folic acid acc. to RDI. 	& Displays energy and nutrient content in the records per food, meal, or day, and data is stored in the server.&	Swedish national food database version 2010-05-05 is used &	Estimated TEE; EI; gram of food; dietary fiber, calcium, iron, vitamin C, vitamin D, and folic acid acc. to RDI &	Not  mentioned	& Estimated with well-suited units (eg, gram, deciliter, tablespoon, teaspoon, and piece).	& Estimated TEE; EI; dietary fiber, calcium, iron, vitamin C, vitamin D, and folic acid &	Validated database \\ \hline
MRP Support App \cite{74}  &
2016 &Only show meal type with no evaluation of the nutrient &	User views  meal and feedback from researcher by clicking the date on the calendar&	No database &	No nutrient and calories shown.
Only type of meal taken &	No nutrients and calories information and Meals are displayed in 5  categories (meal replacements, balanced meals, other meals, allowed snacks, and other snacks)	& No portion size estimation &	No nutrient and calories shown.& No database \\ \hline
Nutricam \cite{75}  &
2011 &User cannot view evaluation  by the dietitian 	& No history. Data directly send to dietitian for analysis	& No&	No diet record&	No	&Take photograph and voice recording explaining the food and/or drink items for consumption. &	Not mentioned &	No database \\ \hline
Easy Diet \cite{57}  &
2017 &	 Calorie and Macro and Micro Nutrients intake&	Type of meal and meals information&	AusBrands 2015) and AusFoods 2015	&Meal type and nutrient Information&	Calories , Micro and Macro nutrient Information of each category of meal &	Common standard measure&	Calories, Macro and Micro Nutrient Components&	Validated database  \\ \hline
Lark  \cite{58}  &
2018 &	Calories intake, Energy Expenditure and messages of health consequences of foods &	Goals Achievement, Meals Calories and nutrient Information 	& USDA database  and traditional recipes &	Meal Type and Calories Information&	Break down of calories and nutrient information per meal &	Not mentioned&	Calories, Micro and Macro Nutrient components &	Validated \\ \hline
Sweetch \cite{63}  &
2018&	Meal Calories and Nutrients Tracking Charts	&Review Summarized History, Meals, calories, Nutrient information , weight changes 	&Not mentioned	&Meal Type, Calories and Nutrient Information&	Not mentioned&	Not mentioned&	Calories &	Not mentioned \\ \hline
Speech2Health \cite{64} &
2018 &	Break down of daily calories and charts for nutrients based on goal&	Select date to review&	USDA database&	Not mentioned&	Break down of daily calories and nutrient components &	Text Processing for mapping Food information to dietary composition	& Calories and nutrients &	Validated database \\ \hline
FoodWiz2 \cite{66} &
2018 &	Track calories intake according to set targets&	Not mentioned For User side but dietician are able to view all meals added from app through web portal 	& UK McCance and Widdowson’s database	& Calories and weight information &	Not mentioned &	Bluetooth scale to estimate portion size&	Calories&	Validated database \\ \hline
Habit App \cite{68} &
2018 &	Tracking calories  intake and steps to achieve weight loss goal	&Able 
To Review Previous history	&
Not mentioned&	Calories&	Calories taken relative to Calories goal &	Not mentioned &	Calories &	Not mentioned \\ \hline
MyFood App \cite{69} &
2018 &	Estimation of Energy, protein and liquid consumption&	Able to Review previous history of Energy, protein and liquid consumption &	Not mentioned	&Energy, protein, liquid consumption, meal type and portion size	&Energy, protein and liquid consumption of each meal type &	Integrated mobile camera for portion estimation &	Calories, protein and liquid consumption	& Not mentioned \\ \hline
My Fitness Pal (Log2Lose) \cite{71} &
2019 &	Breakdown of daily calories and nutrient based on goal&	Select date to review history&	Not mentioned&	Not mentioned&	Breakdown of daily calories and nutrient &	Household measurement &	Calories and nutrients &	Not mentioned \\ 
\hline
\end{longtable} \end{small}
\endgroup
\makeatletter\@ifundefined{TwoColDocument}{}{\twocolumn}\makeatother 
\end{landscape}
\clearpage
\begin{landscape}
\makeatletter\@twocolumnfalse\makeatother
\begingroup
\makeatletter\if@twocolumn\@ifundefined{theposttbl}{\gdef\TwoColDocument{true}\onecolumn\onecolumn}{}\fi\makeatother \setlength\LTcapwidth{\textheight}
\begin{small}
\begin{longtable}{p{2cm}lp{1.5cm}p{3cm}p{3cm}p{3cm}p{3cm}p{4cm}}
\caption{{\textbf{(B) Overview of mHealth apps reporting on the implementation of the dietary components  (Data extracted for RQ4).}} }
\label{table:table9}
\def\arraystretch{1}\\\endfirsthead \hline \noalign{\vskip3pt} \noalign{\textit{Table \thetable\ continued}} \noalign{\vskip 3pt} \endhead \hline \noalign{\vskip3pt} \noalign{\textit{\hfill Continued on next page}} \noalign{\vskip3pt} \endfoot \endlastfoot


\hline
Application Name&	Year&	Include to record total water intake&	Inclusion of nutrient information in the database&	Provide diet guide that cater to individual needs&	Types of nutrients required to be displayed&	Ways to display the summarized energy and nutrient intake \\ \hline
Social POD \cite{16} &
2016&	No&	Calories&	Notifications&	Calories&	Calories per meal \\ \hline
Happy \cite{18} &
2017&	Yes &	Not mentioned&	Messages followed the European Code Against Cancer guidelines&	Not mentioned&	Not mentioned \\ \hline
Lose it \cite{19} &
2017&	Yes &	Calories, fat, sugar&	FB messages for dietary intake&	Calories, fat, sugar&	Calories balanced per meal \\ \hline
mDPP \cite {20}  &
2015&	No&	Calories, SF, sugar&	Tailored feedback was provided&	Calories, SF, sugar	& Calorie diary \\ \hline
Metabolic Diet app \cite{21} &
2016&	Yes	& Energy, carbo, fat, protein, amino acid &	No	&Amino acid, energy, macronutrients&	Nutrient information per serving size, and pie charts showing target and actual intake of specific amino acid and total protein \\ \hline
Diet cam \cite{22} &
2012&	No&	Calories&	Not mentioned	&Calories&	The Album menu organizes all the food items as frequency list, and shows them in a table. \\ \hline
My Fitness Pal \cite{23}&
2017&	Yes	& Calories and nutrients&	Balance calories &	Daily calories and nutrient intake&	Breakdown of daily calories and nutrient in pie chart and table \\ \hline
mpFR \cite{25} &
2010&	No&	Energy and nutrient&	Not mentioned&	Energy and nutrient	&Image analysis with volume estimation and nutrient information \\ \hline
FTA \cite {27} &
2014&	No	&No&	Food habit registration&	Not mentioned	& Food information diary \\ \hline
TreC-LifeStyle \cite{28} & 2017&	Yes	&Calories, carbs, protein, fat	&	Food pyramid and Mediterranean diet guidelines&	Calories, carbs, protein, fat&	Colors of food categories on dashboard, reports of meals and nutrient balance \\ \hline
Snack Track School app \cite{29} &
2016&	No&	Calories&	Flemish guidelines&	Nutritional value&	Size and nutritional content of each snack \\ \hline
My Meal Mate \cite{6,27} &
2013&	Yes	&Calories	&Remaining calories allowance&	Calories&	Remaining calories table and breakdown of calories per meal \\ \hline
FRapp \cite{31} & 2015&	Yes &	Not mentioned&	Real time communication between user \& clinician&	Not mentioned&	Breakdown food images according to meal \\ \hline
iDAT [32] & 2015&	No	&Calories &	Calories left to consume	&Calories&	Calories consumed progress bar, breakdown calories per meal \\ \hline
Metabolic Diet App \cite{20} &	2012&	Yes&	Not mentioned&	Text messages \& dietary feedback	&Not mentioned	&Before and after eating food images table \\ \hline
WIZE \cite{73} &
2015&	No&		Facts, games,\& goals&		List of weekly goals \\ \hline
 ENGAGED \cite{35} &
2012&	No	&Calories \& fat			&	Calories and fat		&Goal thermometers display user goal \& actual amount of calories \& fat gram consumed \\ \hline
Mobile phone apps [36] &
2016 &	No &	Calories&	Personalized recommendation, balance calories&	Calories&	Summary of food intake, a balance of ingested and spent calories \\ \hline
GoCARB \cite{37} &
2016&	Yes&	CHO	&	CHO estimation&	Volume and CHO	& CHO and volume estimation of each food type \\ \hline
CalFit Chi \cite{38} &
2016&	No&	Not mentioned&	Not mentioned&	Standard portion size&	Not mentioned \\ \hline
APPETITT \cite{43} &
2017&	Yes. But actual volume no&	No applicable &	Not mentioned &	No nutrient displayed&	Very simple and to little info shown \\ \hline
DietApp \cite{44}  &
2017&	No&	Only calories info&	Shows food whether harmful or beneficial; user receive personalized dietary suggestions acc. to illnesses they provide in profile.	& Calories, calories burn, balance calories left&	Calories, calories burn from activity, balance calories left all shown in table in one page \\ \hline
e-DIA \cite{47} &
2016&	No&	No	& Not mentioned &	Very minimal. Only gram from user app side&	Simple table only with food item, mealtime and gram \\ \hline
Vegethon \cite{49} &
2017&	No&	Only shows veggies serving	& Not mentioned	& Not mentioned	& Not mentioned \\ \hline
Dietary Intake Assessment \cite{52} &
&	No&	Swedish food database calculated the intake of 49 nutrients&	Energy percentages of each meal in relation to recommended intakes was given. 	& Estimated TEE; EI; gram of food; dietary fiber, calcium, iron, vitamin C, vitamin D, and folic acid &	The user could see details regarding his/her body mass index (BMI); estimated TEE; EI; and intake of macronutrients, fruits and vegetables, dietary fiber, calcium, iron, vitamin C, vitamin D, and folic acid concerning recommended daily intakes. Energy percentages of each meal with recommended intakes were also given. \\ \hline
MRP Support App \cite{74} &
&	No&	No nutrient and calories shown.	&	At the end of each week users received tailored feedback messages, which reflected their progress to date based on their reported compliance to the MRP and their weight loss.	& No nutrient and calories shown.	&No nutrient and calories shown.
Meals shown in 5 simple categories (meal replacements,balanced meals, other meals, allowed snacks and other snacks) for  each mealtime in a table \\ \hline
Nutricam \cite{75} &
2011&	No	&No		&No&	No	&No
\\ \hline
Easy Diet \cite{57} &
2017&	yes&	Calories, Micro and Macro Nutrients	&No	&Calories&	Breakdown of Meal according to meal time \\ \hline
Lark \cite{58} &
2018&	No	&Calories, Micro and Macro Nutrient Information	&	Notifications, messages on behavior and outcome of behavior&	Calories	&Goals, Break down of Meal according to Meal Type, Outcomes of various Foods \\ \hline
Sweetch \cite{63} &
2018&	No&	Calories, Micro And Macro Nutrient Information	&Notifications&	Calories&	Calories Consumed and Total Calories \\ \hline
Speech2Health \cite{64} &
2018&	Yes&	Calories and Nutrients Information &	Balance Calories&	Daily calories and nutrient intake	&Breakdown of daily calories and nutrient in pie chart and table \\ \hline
FoodWiz2 \cite{66} &
2018&	No&	Calories	&	Personalized Feedback text messages for motivation and help &	Calories&	Breakdown of daily Calories \\ \hline 
Habit App \cite{68} &
2018&	No	&Calories	&Problem Solving process for weight loss and breaking bad habits&	Calories&	Calories taken by user relative to calories goal \\ \hline
My Food App \cite{69}&
2018&	Yes&	Energy and protein& 	Not mentioned&	Calories and protein &	Breakdown of calories and protein intake according to meal type  \\ \hline
My Fitness Pal (Log2Lose) \cite{71} &
2019&Yes&	Calories and nutrients&	User Receives weekly Feedback on goals and their financial incentives.& Daily calories and nutrients intake &		Breakdown of daily calories and nutrient \\ 
\hline
\end{longtable}
\end{small}
\endgroup
\makeatletter\@ifundefined{TwoColDocument}{}{\twocolumn}\makeatother 
\end{landscape}
\end{document}